\documentclass[10pt, pre, amsmath, twocolumn, showpacs, superscriptaddress]{revtex4-1}
\usepackage{graphicx}
\usepackage{amsfonts}
\usepackage{bm}
\usepackage{multirow}
\usepackage{xcolor}
\usepackage{tikz}
\usetikzlibrary{shapes,shapes.multipart}

\definecolor{dgreen}{rgb}{0,0.7,0}

\usepackage[normalem]{ulem}

\begin{document}
\title{Run-and-Tumble particle in one-dimensional confining potential: Steady state, relaxation and first-passage properties}

\author{Abhishek Dhar}
\address{International Centre for Theoretical Sciences, TIFR, Bangalore 560089, India}

\author{Anupam Kundu}
\address{International Centre for Theoretical Sciences, TIFR, Bangalore 560089, India}

\author{ Satya N. Majumdar}
\address{LPTMS, CNRS, Univ. Paris-Sud, Universit\'{e} Paris-Saclay, 91405 Orsay, France}

\author{Sanjib Sabhapandit}
\address{Raman Research Institute, Bangalore 560080, India}

\author{Gr\'egory Schehr}
\address{LPTMS, CNRS, Univ. Paris-Sud, Universit\'{e} Paris-Saclay, 91405 Orsay, France}
\date{\today}

\begin{abstract}
\noindent 
{We study the dynamics of a one-dimensional run and tumble particle subjected to confining potentials of the type $V(x) = \alpha \, |x|^p$, with $p>0$. The noise that drives the particle dynamics is telegraphic and alternates between $\pm 1$ values. We show that the stationary probability density $P(x)$ has a rich behavior in the $(p, \alpha)$-plane. For $p>1$, the distribution has a finite support in $[x_-,x_+]$ and there is a critical line $\alpha_c(p)$ that separates an active-like phase for $\alpha > \alpha_c(p)$ where $P(x)$ diverges at $x_\pm$, from a passive-like phase for $\alpha < \alpha_c(p)$ where $P(x)$ 
vanishes at $x_\pm$. For $p<1$, the stationary density $P(x)$ collapses to a delta function at the origin, $P(x) = \delta(x)$. In the marginal case $p=1$, we show that, for $\alpha < \alpha_c$, the stationary density $P(x)$ is a symmetric exponential, while for $\alpha > \alpha_c$, it again is a delta function $P(x) = \delta(x)$. For the harmonic  case $p=2$, we obtain exactly the full time-dependent distribution $P(x,t)$, that allows us to study how the system relaxes to its stationary state. In addition, for this $p=2$ case, we also study analytically the full distribution of the first-passage time to the origin. Numerical simulations are in complete agreement with our analytical predictions.}
\end{abstract}

\maketitle

\section{Introduction}
\label{intro}

There has been a surge of interest in understanding the statics and the dynamics of a class of
non-equilibrium systems, commonly called ``active matter''. Such systems appear in a variety of contexts ranging from self-catalytic colloids \cite{Fodor17}, living cells, biological processes, in the growth of microbial colonies \cite{Magistris15}, in the physics of self-motile systems \cite{Magistris15},  active gels \cite{Sriram10}, motility induced phase transition \cite{Cates15} to flocking of birds \cite{Toner05}. While the collective properties of such systems are of great interest \cite{Fodor17,bechinger_active_2016,cates_motility-induced_2015}, they exhibit a rich static and dynamical behaviour even 
at a single-particle level. This is due to the fact that the single-particle stochastic dynamics is typically non-Brownian: the 
effective noise in the Langevin equation of an active particle is typically ``coloured'', i.e., the autocorrelation of the noise
decays exponentially $\sim e^{-t/\tau}$ with a finite correlation time $\tau$~\cite{footnote}. The limit $\tau \to 0$ would correspond to the 
Brownian motion, while the $\tau \to \infty$ limit corresponds to purely ballistic motion.

\begin{figure}[ht]
\centering
 \includegraphics[width = 0.8\linewidth]{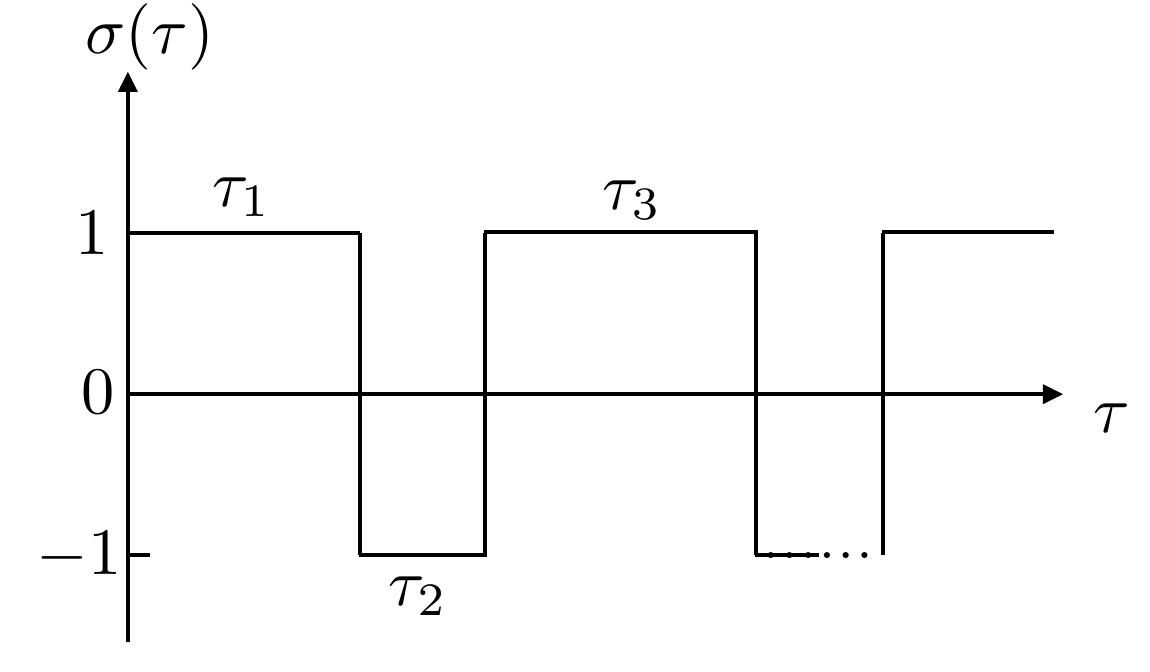} 
 \caption{A typical realisation of the telegraphic noise $\sigma(\tau)$,  which {alternates its value between $\pm1$}  at {random} times $\tau_i$'s. The time intervals between two successive reversals of the sign of $\sigma(\tau)$ are statistically independent and each is distributed exponentially as $p(\tau)=\gamma e^{-\gamma \tau}$.}
\label{fig:sigma-t}
\end{figure}

The most common example of such an active particle dynamics is the well known ``persistent Brownian 
motion'', renamed recently as the ``run and tumble particle'' (RTP). In one-dimension, the overdamped 
stochastic dynamics of an RTP, starting at $x=x_0$, is governed by the Langevin equation 
\begin{equation}
\frac{d x}{d t} = v_0\,\sigma(t), \label{RTP_free}
\end{equation}
where the noise $\eta(t) = v_0\,\sigma(t)$ is a dichotomous or  
telegraphic noise which takes values $\pm v_0$ and flips its sign with a constant rate $\gamma$, as shown in Fig~\ref{fig:sigma-t}. The auto-correlation function of this telegraphic noise is given by 
\begin{eqnarray}\label{autocorrel}
\langle \eta(t_1) \eta(t_2) \rangle = v_0^2\,e^{-2 \gamma\,|t_1-t_2|}  \;.
\end{eqnarray}
In the limit $\gamma \to \infty$, $v_0 \to \infty$ keeping the ratio $v_0^2/(2\gamma)=D$ fixed, the noise $\eta(t)$ becomes a white noise
with correlator $\langle \eta(t_1) \eta(t_2) \rangle = 2 D\,\delta(t_1-t_2)$ and consequently $x(t)$ represents the position of a Brownian particle. 
{Indeed, this problem of  a single particle driven by telegraphic noise is generally known as ``persistent random walk" and 
has been studied extensively. Several properties of this walk in one dimension, with or without boundaries (absorbing or reflecting),  have been studied~\cite{Othmer, Masoliver92, Leydolt93, Bicout97, Weiss02,Martens12, Sim15, Demaerel18, Malakar2018}.
}

What happens if this RTP is subjected to an external potential $V(x)$? The corresponding Langevin equation now reads
\begin{equation}
\frac{d x}{d t} = f(x) + v_0\,\sigma(t), \label{RTP}
\end{equation}
where $f(x) = -V'(x)$ is the external force. Once again, in the Brownian limit, when $\gamma \to \infty$, $v_0 \to \infty$ keeping the ratio $v_0^2/(2\gamma)=D$ fixed, the probability distribution $P(x,t)$ converges at long times to the equilibrium Boltzmann-Gibbs distribution $P(x) \propto e^{- V(x)/D}$. However, when the correlation time of the noise $\tau = 1/(2 \gamma)$ is finite, i.e. the noise has a finite memory, the stationary state $P(x)$ is nontrivial and is typically non-Boltzmann. While an explicit formula is known for this $P(x)$, originally derived in the quantum optics literature~\cite{Horsthemke84, Klyatskin78a,Klyatskin78b,Lefever80,Hanggi95} and re-derived more recently in the active-matter literature~(see for instance the Appendix in~\cite{Cates_Nature}), the structure of the stationary state has not been analysed in detail for different types of potentials. One of the goals of this paper is to study this stationary state for a class of confining potentials of the type $V(x) = \alpha\,|x|^p$ with $p>0$ and $\alpha > 0$. Indeed, keeping $\gamma$ fixed, and varying $p$ and $\alpha$, we uncover an interesting phase diagram in the $(p,\alpha)$ plane. For $p>1$, we show that 
while the stationary solution has a finite support for any $\alpha$, there is a critical line $\alpha_c(p)$     
\begin{eqnarray}\label{gammac_intro}
{\alpha=\alpha_c(p) =  \left(\frac{v_0}{p}\right)^{2-p}\left(\frac{\gamma}{p(p-1)} \right) ^{p-1}},
\end{eqnarray}
such that for $\alpha < \alpha_c(p)$ the stationary state distribution vanishes at the edges, while for $\alpha >\alpha_c(p)$ the distribution diverges at the edges.  
Thus, while for $\alpha > \alpha_c(p)$ the particle behaves more like an ``active'' one, it behaves more as a ``passive'' particle for $\alpha < \alpha_c(p)$. 
 We call this transition from the active  to the passive phase across $\alpha=\alpha_c(p)$ a ``shape transition'' of the stationary profile. 
The behaviour of $P(x)$ around $x=0$ depends on $p$ and is also interesting. We find that for $1<p < 2$, the function $P(x)$ near the origin is convex {(i.e., there is a local minimum at $x=0$)} while for $p>2$ it is concave {(i.e., there is a local maximum at the origin)}, irrespective of whether $\alpha > \alpha_c$ or $\alpha < \alpha_c$. {A weak fingerprint} of activeness is thus manifest, for $1<p <2$, even in the passive phase, leading to bimodal distributions  (see for instance Fig.~\ref{fig6}).
For $p<1$, we show that the RTP collapses to the origin at long times, leading to trivial stationary state $P(x) = \delta(x)$. The case $p=1$ turns out to be the marginal case. In this case there is a nontrivial stationary state only for $\alpha < v_0$, while the stationary distribution is again $P(x) = \delta(x)$ for $\alpha > v_0$. These results lead to the phase diagram shown in Fig \ref{Fig_Ph_Diag}.

In the presence of the external confining potential $V(x)$, it is also equally interesting to study the dynamics of the system away from the stationary state, i.e., how the system relaxes to the stationary state at long times. For a simple hard-box potential, where the RTP is confined inside a finite box, e.g. when $V(x) = 0$ for $x \in [-1,1]$ while $V(x) = +\infty$ outside the box, this relaxation to the stationary state was recently studied in Ref. \cite{Malakar2018}. This case corresponds to the $p \to \infty$ limit of the potential $V(x) = \alpha \, |x|^p$. 
It is thus interesting to study the relaxational dynamics for a finite value of $p$. In this paper, we  
provide a detailed study of the relaxational dynamics for the case $p=2$, i.e. the harmonic potential $V(x) = \alpha \, x^2$ with $\alpha=\mu/2$. Interestingly, our results show that, if the particle starts exactly at the origin, i.e., $x_0=0$, the inverse relaxation time $\lambda_0$ characterizing the exponential decay of $P(x,t) - P(x) \sim e^{-\lambda_0\,t}$ to its stationary distribution undergoes a transition at $\alpha = \gamma$: for $\alpha > \gamma$ (passive phase) $\lambda_0 = \mu + 2 \gamma$, while for $\alpha < \gamma$ (active phase) we show that $\lambda_0 = 2 \mu$ (see Fig. \ref{Fig_lambda0} ). For a generic starting point $x_0  \neq 0$, the inverse relaxation time is always given by $\mu$. Thus if one wants to detect the signature of active to passive transition in the relaxational dynamics, one needs to start at the special initial position $x_0 = 0$.

Another related interesting observable is the first-passage probability $F_{\rm fp}(t,x_0)$ which denotes the probability density that the RTP, starting at $x_0>0$, crosses the origin for the first time at time $t$. For arbitrary confining $V(x)$, there exist extensive results in the literature on the {\it mean} first-passage time~\cite{Lindenberg,Weiss_Lind}. However, there is hardly any result on the full distribution of the first-passage time. Only very recently exact results were obtained for a free RTP, i.e. for $f(x) = 0$, as well as for the box potential \cite{Malakar2018}. In this paper, we present exact results for $F_{\rm fp}(t,x_0)$ for two cases $p=2$ and $p=1$.

The rest of the paper is organised as follows. In Section \ref{steady-state}, we provide a derivation of the steady state probability density for 
arbitrary confining potential $V(x)$ and then discuss the form of the stationary distribution for potentials of the form $V(x) = \alpha \, |x|^p$, with $p>0$.
We show that there is a shape transition in the $(p,\alpha)$-plane from an active profile to a passive profile across a critical line $\alpha_c(p)$. The resulting phase diagram is shown in Fig.~\ref{Fig_Ph_Diag}. In Section \ref{relax-harmonic}, we discuss the relaxation to the steady state for the harmonic potential $p=2$. First-passage properties for this harmonic case are discussed in Section \ref{FP}. We conclude in Section \ref{sec:conclusion} with a summary and open questions. Some details have been relegated to two Appendices. In Appendix \ref{appendixA} we provide some details of the calculations of the first-passage properties for the harmonic case $p=2$.

\section{Steady state distribution}
\label{steady-state}

\begin{figure}
\includegraphics[width = \linewidth]{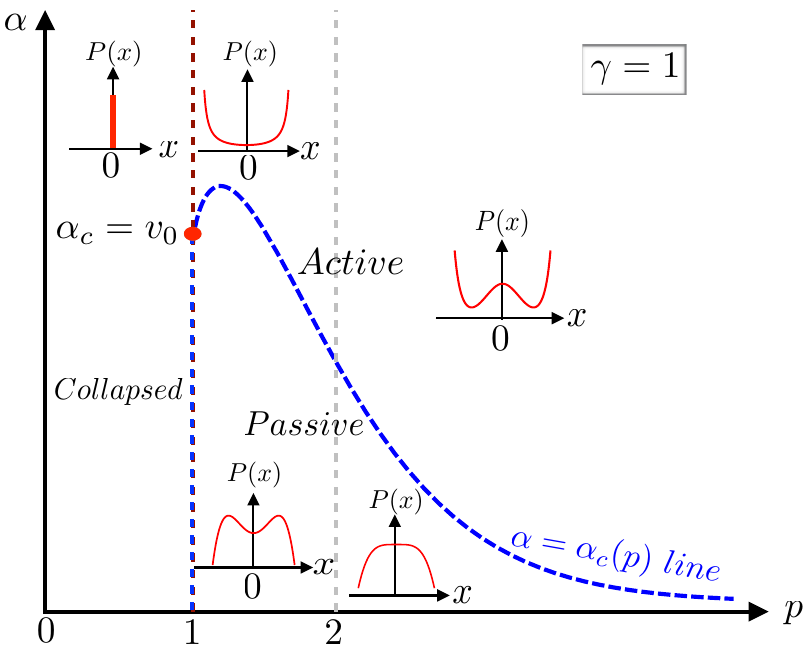}
\caption{Phase diagram in the $(p, \alpha)$ plane. For $p>1$, the blue dashed line $\alpha_c(p)$ separates an active-like
phase for $\alpha > \alpha_c(p)$ and a passive-like phase for $\alpha < \alpha_c(p)$. {The stationary solution $P(x)$ is  divergent at the edges in the active-like phase while it goes to zero in the passive-like phase.} Exactly on the marginal line $p=1$, there is a critical value $\alpha_c = v_0$: for $\alpha < \alpha_c = v_0$, $P(x)$ has a symmetric exponential decay while, for $\alpha \geq \alpha_c = v_0$, the stationary state collapses to a delta function at the origin, i.e., $P(x) = \delta(x)$. For $p<1$, for all $\alpha >0$, the stationary state is again given by a delta function, $P(x) =~\delta(x)$. To obtain this phase diagram we have chosen $\gamma=1$. However, for other 
values of $\gamma$ the phase diagram is qualitatively the same.}\label{Fig_Ph_Diag}
\end{figure}

We consider the Langevin equation (\ref{RTP}) for a particle moving in an arbitrary confining potential $V(x)$ with the force $f(x) = - V'(x)$ and
subjected to the telegraphic noise $v_0\,\sigma(t)$. {We focus on a particular family of  confining potentials, of the type 
$V(x) = \alpha \,|x|^p$ with $p>1$, such that the system reaches a normalizable stationary state at long times.} This steady state distribution is actually known for generic confining potential $V(x)$. It was initially derived in the quantum optics literature~\cite{Horsthemke84, Klyatskin78a,Klyatskin78b,Lefever80}, later to study the role of colored noise in dynamical systems \cite{Hanggi95} and more recently it has been re-derived in the context of active particles systems~\cite{Cates_Nature}. Even though an explicit formula existed for the stationary state, the physical consequences of this formula has not been analysed in detail in the literature, to the best of our knowledge {[see, however, a recent article~\cite{Sevilla18} that discusses the behavior of the stationary distribution for a general class of potentials, including $V(x)= \alpha |x|^p$, but only for integer $p$].
Our purpose in this section is to focus on this particular family of potentials $V(x) = \alpha \,|x|^p$ with arbitrary $p>1$, and derive the detailed behavior of the shape of the stationary distribution in the full $(p,\alpha)$ plane.  We show that the existence of a noval shape transition in the $(p,\alpha)$ plane (for  fixed $\gamma$) across a crossover line $\alpha_c(p)$ for $p>1$  that separates the active and the passive phases.
The exact formula for $\alpha_c(p)$ for $p>1$ is given in Eq.~(\ref{gammac}). For the full phase diagram in the $(p,\alpha)$ plane, see
Fig. \ref{Fig_Ph_Diag}. 
 }

We start by deriving the Fokker-Planck equation associated to the Langevin equation (\ref{RTP}). Since the telegraphic noise $v_0\,\sigma(t)$ in (\ref{RTP}) takes only two values $\pm v_0$, it is natural to define $P_+(x,t)$ and $P_-(x,t)$ denoting the probability densities for the particle to be at position $x$ at time $t$ with velocity $+v_0$ and $-v_0$, respectively. Clearly, the main object of interest is the full probability density $P(x,t) = P_+(x,t) + P_-(x,t)$. It is quite easy to see that these two densities $P_+(x,t)$ and $P_-(x,t)$ evolve according to
\begin{eqnarray}
\label{eqn:P1}
 { \frac{\partial P_{+} }{\partial t}} &=& -\frac{\partial}{\partial x} \left[(f(x)+v_0)P_{+}\right] - \gamma \, P_{+} +  \gamma P_{-} \, ,  \\
{ \frac{\partial P_{-}}{\partial t}}  &=& -\frac{\partial}{\partial x} \left[(f(x)- v_0)P_{-}\right] + \gamma \, P_{+} -  \gamma P_{-} \, . 
\label{eqn:P2}
\end{eqnarray}
where the first term in both the equation appear from the drift term $f(x) \pm v_0$ in the Langevin Eq.~\eqref{RTP}, while the remaining terms appear due to the stochastic change in the direction of the velocity with rate $\gamma$.  For sufficiently confining potentials, we expect the system to reach a stationary state which is obtained by setting the time derivative to be zero on the left hand side in Eqs. (\ref{eqn:P1}) and (\ref{eqn:P2}). We denote the stationary distributions $P_\pm(x)$ (dropping the $t$-dependence) which then satisfy a pair of coupled first-order differential equations
\begin{eqnarray}
\label{eqn:p+}
 \frac{d}{d x} \left[(f(x)+v_0)P_{+}\right] + \gamma \, P_{+} -  \gamma P_{-}&=&0 ,  \\
\frac{d}{d x} \left[(f(x)- v_0)P_{-}\right] - \gamma \, P_{+} + \gamma P_{-} &=&0 \;.  \label{eqn:p-} 
\end{eqnarray}

To solve these differential equations, we need to specify the boundary conditions for $P_+(x)$ and $P_-(x)$. For that, we first need to know where the boundaries are. In fact, it turns out that in the steady state, for sufficiently confining potentials, these active particles accumulate over a finite region of space around the minimum of the confining potential (assuming that it has a single minimum, such as in a harmonic potential). This can be easily seen by examining the Langevin equation (\ref{RTP}). We first note that the right hand side  of Eq. (\ref{RTP}) has two fixed points $x_{\pm}$ such that 
\begin{eqnarray}\label{def_fp}
f(x_-) = v_0 \;, \; f(x_+) = -v_0 \;,\; {\rm with} \;\; x_+ > x_- \;.
\end{eqnarray}

\begin{figure}
\includegraphics[width = \linewidth]{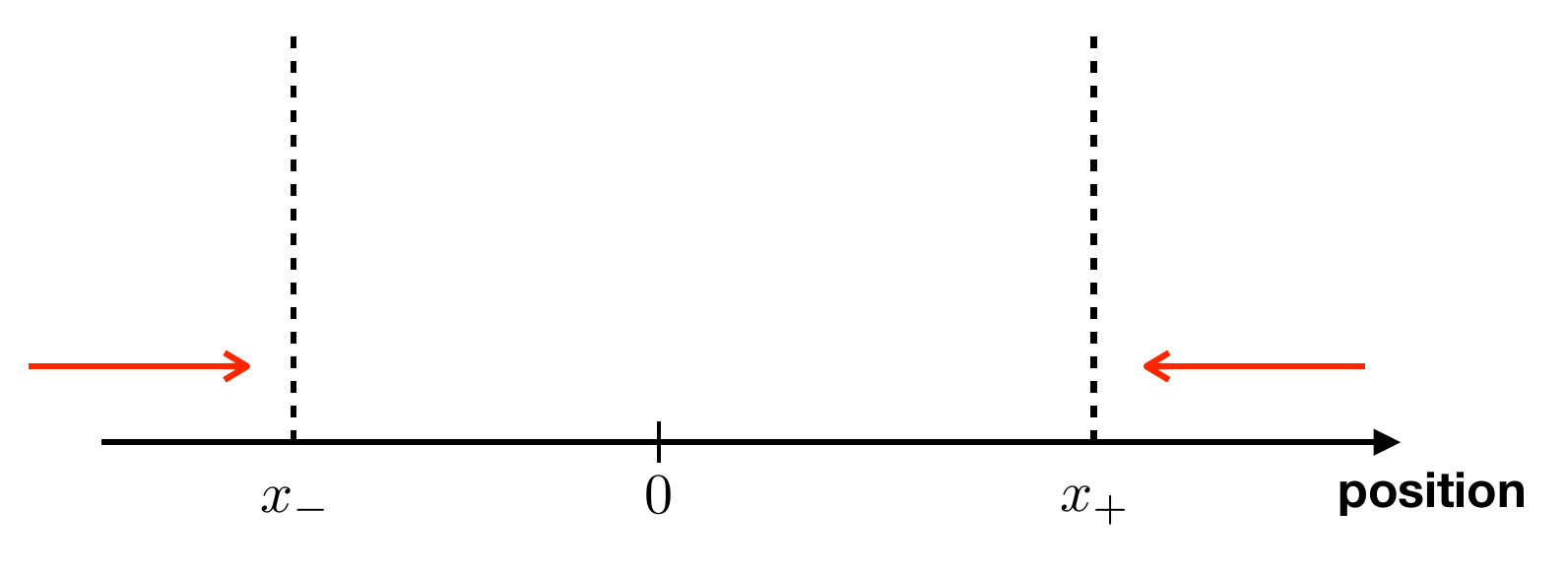}
\caption{Illustration of the confinement of the particle within the finite box $[x_-, x_+]$, where $f(x_{\pm}) = \mp v_0$ [see Eq. (\ref{def_fp})]. The red arrows represent the direction of the drift felt by
the particle outside of the box [see Eq. (\ref{RTP})].}\label{Fig:confined}
\end{figure}
For example, for a harmonic potential $V(x) = \mu\,x^2/2$, $f(x) = - \mu \,x$ and hence $x_{\pm} = \pm v_0/\mu$. Consider a particle with position $x>x_+$. Then the drift is always towards the center of the trap, no matter what the sign of $\sigma(t)$ is, as long as $x>x_+$ (see Fig. \ref{Fig:confined}). Thus, even if the particle starts with initial position $x_0>x_+$, it will eventually arrive at a position $x<x_+$. Similarly, for any particle with position $x \leq x_-$, the drift is always towards the center, irrespective of the sign of $\sigma(t)$. Hence, eventually, the position of the particle will be larger than $x_-$, even if it starts to the left of $x_-$ (see Fig. \ref{Fig:confined}). Thus it is clear that, in the steady state, the probability distribution will have a finite support $[x_-,x_+]$ where it is non-zero, while it vanishes outside this box (see Fig. \ref{Fig:confined}). Therefore, the steady state equations in (\ref{eqn:p+}) and (\ref{eqn:p-}) are valid for $x \in [x_-,x_+]$ and we need to specify the boundary conditions at $x=x_{\pm}$. To find these boundary conditions, we note that if the particle is at $x_+$ with a negative velocity, it always moves to the left of $x_+$. Therefore, in the stationary state, there can not be any particle at $x_+$ with a negative velocity, indicating that 
\begin{eqnarray}
P_-(x_+) = 0 \;. \label{bc_p+}
\end{eqnarray}
Note that $P_+(x)$ remains unspecified at $x=x_+$. A similar argument at $x_-$ shows that 
\begin{eqnarray}
P_+(x_-) = 0 \;, \label{bc_p-}
\end{eqnarray}
while $P_-(x)$ remains unspecified at $x=x_-$. With this pair of boundary conditions (\ref{bc_p+}) and (\ref{bc_p-}), the differential equations in Eqs. (\ref{eqn:p+}) and (\ref{eqn:p-}) admit a unique solution that we determine below.

 \begin{figure}[t]
\centering
 \includegraphics[width=0.9\linewidth]{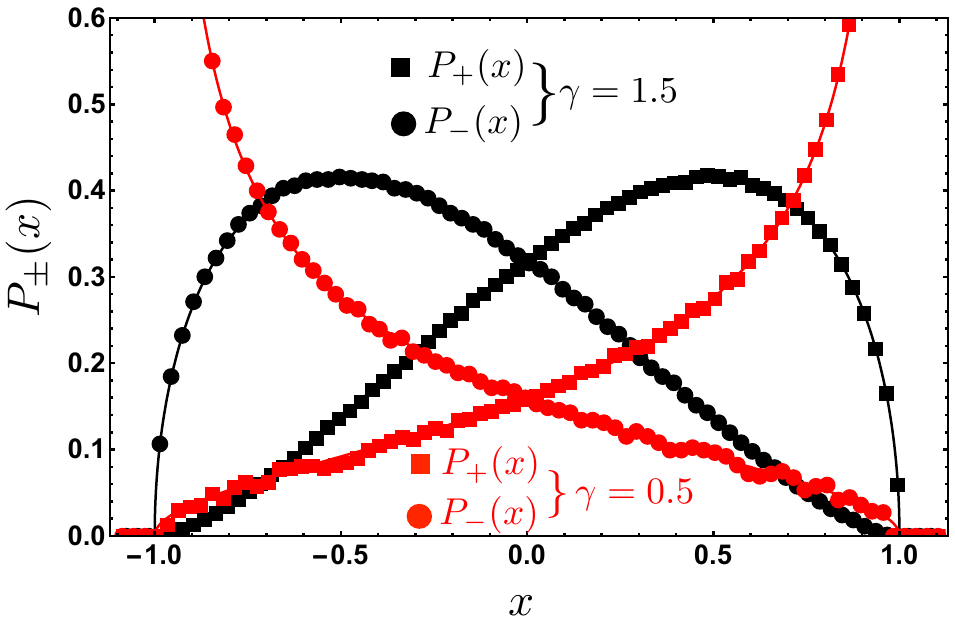} 
 \caption{{Numerical verification of the theoretical results of the distributions $P_\pm(x)$  in \eqref{P_pm-harm} for the harmonic potential $V(x) = \alpha\,x^2 = \mu\,x^2/2$, setting $v_0=1$ and $\mu = 1$, i.e. $\alpha = 1/2$. In this case $x_{+} = +1$ and $x_-=-1$. The filled black squares correspond to $P_+(x)$ whereas the filled black circles correspond to $P_-(x)$ obtained from the simulation for $\gamma=1.5$. The red points correspond to another value of $\gamma = 0.5$. The solid lines are obtained from the theoretical expression in Eq. (\ref{P_pm-harm}), respectively for $\gamma = 1.5$ and $\gamma=0.5$. We observe that, as in the case of $P(x)$, the distributions $P_\pm(x)$ also undergo a shape transition as one varies $\gamma$ across $\gamma_c=2\alpha= 1$. Thus $\gamma = 1.5 > \gamma_c = 1$ corresponds to the passive-like regime, while $\gamma = 0.5 < \gamma_c=1$ corresponds to the active-like regime. Moreover in the active-like regime ($\gamma < \gamma_c$) the distributions $P_\pm(x)$ have diverging (integrable) behavior at the edges $x=x_\pm$ whereas they satisfy the boundary conditions in Eqs. \eqref{bc_p+} and \eqref{bc_p-} on the other edges $x=x_\mp $ respectively.}}
\label{fig:p-pm-harm}
\end{figure}

To proceed, we add and subtract these two equations (\ref{eqn:p+}) and (\ref{eqn:p-}). Defining 
\begin{align}
&P(x)=P_+(x)+P_-(x), \label{P-m} \\
&Q(x)=P_+(x)-P_-(x) \;,\label{Q-m}
\end{align}
they satisfy, respectively, the equations
\begin{align}
&\frac{d}{d x} \left[f(x)\, P+v_0\,Q\right]=0, \label{eq-1}\\
 &\frac{d}{d x} \left[f(x)\,Q+v_0\,P\right] + 2 \gamma Q=0 \;. \label{eq-2}
\end{align}
Eq. \eqref{eq-1} implies 
\begin{align}
 &\left[f(x)\,P+v_0\,Q\right]=C, \label{eq-C}
\end{align}
where $C$ is a constant. To determine this constant, we now invoke the boundary conditions (\ref{bc_p+}) and (\ref{bc_p-}). Clearly, Eq. (\ref{eq-C}) holds for all $x \in [x_-,x_+]$. Setting $x=x_-$ we get
\begin{equation}\label{eq-C2}
\left[f(x_-) + v_0\right] P_+(x_-) + \left[f(x_-) - v_0 \right] P_-(x_-) = C \;.
\end{equation}
Using $f(x_-) = v_0$ from Eq. (\ref{def_fp}), and the boundary condition $P_+(x_-) = 0$ from Eq. (\ref{bc_p+}), we get $C=0$. Hence from Eq. (\ref{eq-C}), we get
\begin{eqnarray}\label{expr_Q}
Q(x) = - \frac{1}{v_0}\,f(x) P(x) \;.
\end{eqnarray}
We now substitute this equation in Eq. (\ref{eq-2}) and obtain
\begin{eqnarray}
\frac{d}{d x} \left[ (v_0^2 - f^2(x))P(x) \right] -2\gamma f(x)\, P(x) =0 \;. ~~~\label{p_ss-1}
\end{eqnarray}
Solving this equation, we get
\begin{equation}
P(x) = \frac{A}{v_0^2 -f^2(x)} \exp \left( 2 \gamma \int_0^x dy \frac{f(y)}{v_0^2 -f^2(y)}\right), \label{pss-gen}
\end{equation}
valid for $x_- \leq x \leq x_+$. The overall constant $A$ in (\ref{pss-gen}) is determined from the normalisation $\int_{x_-}^{x_+} P(x)\, dx =1$. This provides the general steady state solution for arbitrary $f(x)$, provided $P(x)$ is normalisable. Below, we analyse this formula for the stationary distribution (\ref{pss-gen}) for confining potentials of the type $V(x) = \alpha \, |x|^p$, i.e., with $f(x) = -\alpha \, p~{\rm sgn}(x)\,|x|^{p-1}$. We will see below that the situation is quite different for $p>1$ and $p<1$, including the marginal case $p=1$. We first start with the $p=2$ case, which corresponds to a harmonic potential $V(x) = \alpha x^2$.

 \begin{figure}[t]
\centering
 \includegraphics[width = \linewidth]{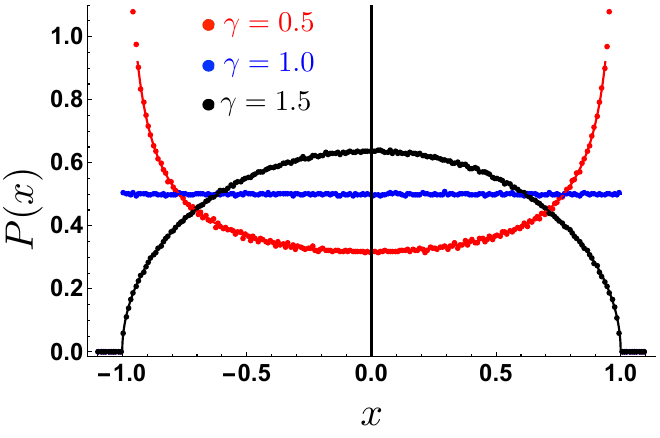} 
 \caption{Numerical verification of the theoretical result of the steady state distribution in \eqref{harm-ss} for the harmonic potential $V(x) = \alpha\,x^2 = \mu\,x^2/2$, setting $v_0=1$ and $\mu = 1$, i.e. $\alpha = 1/2$. 
 In this case, the shape-transition can be accessed by varying the parameter $\gamma$ (as discussed in the text), for fixed $\alpha = 1/2$. The critical value of $\gamma$ is $\gamma_c = 2 \alpha=1$. For $\gamma<\gamma_c = 1$, the theory predicts an active-like phase where $P(x)$ is U-shaped, while for $\gamma>\gamma_c = 1$, a passive-like phase with a bell-shaped $P(x)$ (more appropriately an inverted U-shaped $P(x)$). Exactly at the critical point $\gamma = \gamma_c =1$, one would obtain a flat distribution $P(x)$ as predicted in \eqref{harm-ss} with $\phi = 0$. We performed simulations for three different values of $\gamma = 1/2$ (red), $1$ (blue) and $3/2$ (black) to probe these three different cases. The agreement with the theoretical predictions in \eqref{harm-ss} is excellent.}
\label{fig:pss-harm}
\end{figure}

\vspace*{0.3cm}
\noindent{\it Harmonic potential ($p=2$)}. To analyse the implications of this solution in Eq. (\ref{pss-gen}), let us first focus on the specific example of a harmonic potential $V(x) = \alpha x^2$ and for using the standard notations of the harmonic oscillator we set $\alpha =\mu/2$ where $\mu$ is the standard spring constant. In this case, using $f(x) = - \mu x$, and $x_{\pm} = \pm v_0/\mu$, one gets from Eq. (\ref{pss-gen})
\begin{equation}
P(x) = \frac{2}{4^{\gamma/\mu}B(\gamma/\mu,\gamma/\mu)}
~\frac{\mu}{v_0} \left[ 1- \left(\frac{\mu x}{v_0}\right)^2\right]^{\phi}, \label{harm-ss}
\end{equation}
where $B(\alpha,\beta)$ is the beta function and the exponent 
\begin{eqnarray}\label{phi_exp}
\phi = \frac{\gamma}{\mu} - 1 = \frac{\gamma}{2\alpha} - 1\;,
\end{eqnarray}
where we used $\alpha = \mu/2$. Clearly, this solution is symmetric around $x=0$ and has a finite support over $-v_0/\mu \leq x \leq v_0/\mu$. 
We have verified this prediction in Eq. (\ref{harm-ss}) numerically for two different values of $\gamma$ (see Fig.~\ref{fig:pss-harm}). Interestingly, the shape of the solution in Eq. (\ref{harm-ss}) depends on the exponent $\phi$. If $\phi>0$, the solution vanishes at the two ends $\pm v_0/\mu$ of the support, while for $\phi<0$, the solution diverges at the two ends. This shape transition occurs at $\phi=0$ which corresponds to a critical value $\gamma_c=\mu = 2 \alpha$. Indeed, the solution is {concave-shaped} for $\gamma>\gamma_c$ and {convex-shaped} for $\gamma<\gamma_c$. To appreciate this shape transition, it is instructive to compare this solution for the active particle to that of a passive particle described by the Langevin equation (\ref{RTP}) with the same force $f(x) = - \mu x$, but driven by a delta-correlated white noise. In this passive case, Eq. (\ref{RTP}) just describes an Ornstein-Uhlenbeck (OU) process, whose steady state has the Boltzmann distribution 
\begin{eqnarray}\label{ss-OU}
P_{\rm OU}(x) = \sqrt{\frac{\mu}{2\pi D}} e^{-\frac{\mu}{2D}\, x^2} \;,
 \end{eqnarray}    
valid over the full space. It is easy to check that the active solution in Eq. (\ref{harm-ss}) actually approaches the Boltzmann distribution in (\ref{ss-OU}) in the limit $\gamma \to \infty$, $v_0 \to \infty$ but keeping the ratio $v_0^2/\gamma = 2\,D$ fixed. Indeed, in the $\gamma \to \infty$ limit, the telegraphic noise approaches the delta-correlated noise and naturally the two stationary solutions also coincide. However, for finite $\gamma$, the active noise has an exponentially decaying memory and this leads to highly non-Boltzmann distributions in Eq. (\ref{harm-ss}). While for $\gamma > \gamma_c$, the stationary distribution $P(x)$ still retains a single peak structure (as in the $\gamma \to \infty$ Boltzmann case), it undergoes a ``shape transition'' at $\gamma = \gamma_c$. For $\gamma < \gamma_c$, the stationary distribution has a double-peaked structure (with diverging peaks at $x = \pm v_0/\mu$), with a minimum at the center of the trap at $x=0$. Thus, the single-peak structure is a signature of a passive phase, while the double-peak structure signifies an active phase. By reducing $\gamma$, one crosses over from this passive phase ($\gamma > \gamma_c$) to an active phase ($\gamma < \gamma_c$). Alternatively, this shape transition can also be viewed as a function of $\alpha$ (for fixed $\gamma$). There is a critical value $\alpha_c = \gamma/2$ at which the exponent $\phi$ in Eq.~(\ref{phi_exp}) changes sign. The active phase $\gamma<\gamma_c$ (where $\phi<0$) corresponds to $\alpha > \alpha_c$ and the passive phase $\gamma>\gamma_c$ (where $\phi>0$) corresponds to $\alpha<\alpha_c$.

Finally, we note that Eq. (\ref{harm-ss}) gives the expression for the total probability distribution $P(x) = P_+(x) + P_-(x)$. It is instructive to compute $P_+(x)$ ans $P_-(x)$ separately also, as they have rather different behaviours because the boundary conditions they satisfy are quite different [see Eqs. (\ref{bc_p+}) and (\ref{bc_p-})]. Indeed from Eq. (\ref{expr_Q}), for $f(x) = - \mu x$, we have 
\begin{eqnarray}\label{diff}
P_+(x) - P_-(x) = Q(x) = \frac{\mu\,x}{v_0} P(x) \;.
\end{eqnarray} 
This, along with $P(x) = P_+(x) + P_-(x)$, allows to express $P_+(x)$ and $P_-(x)$ in terms of $P(x)$ in Eq. (\ref{harm-ss}). This gives
\begin{align}
P_\pm(x)= \frac{A}{2} \left(1\pm\frac{\mu x}{v_0}\right)^{\frac{\gamma}{\mu}} \left(1\mp\frac{\mu x}{v_0}\right)^{\frac{\gamma}{\mu}-1}, \label{P_pm-harm} 
\end{align}
where $A$ is the normalization constant appearing in Eq. (\ref{harm-ss}), fixed by $\int_{x_-}^{x_+} [P_+(x)+ P_-(x)] \, dx = 1$. Note that the these distributions satisfy the boundary conditions in Eqs. \eqref{bc_p+} and \eqref{bc_p-} respectively. In fig. \ref{fig:p-pm-harm} we verify these expressions of $P_\pm(x)$ numerically and observe a nice agreement.

  \begin{figure}[t]
\centering
 \includegraphics[width = \linewidth]{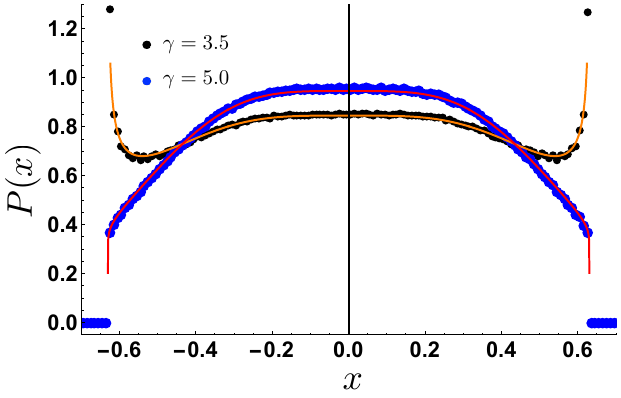} 
 \caption{Numerical verification of the theoretical result of the steady state distribution in \eqref{exact_P1} for $V(x)=\alpha \,x^4$. We set the parameter values $\alpha = 1$ and $v_0 = 1$ and vary $\gamma$ to probe the shape-transition, as done for the harmonic case in Fig. \ref{fig:pss-harm}. In this case, the critical value of $\gamma_c$, from Eq. (\ref{gammac}) with $p=4$, is given by $\gamma_c = 12\times 4^{-2/3} \approx 4.7622$. We considered two representative values of $\gamma = 3.5 < \gamma_c$ (black) and $\gamma = 5 > \gamma_c$ (blue) corresponding respectively to the active and passive-like phases. We find an excellent agreement between the numerical simulations and the theoretical predictions in Eq. (\ref{exact_P1}).}
\label{fig:pss-xpw4}
\end{figure}

\vspace*{0.3cm}
\noindent{\it Anharmonic confining potentials $V(x) = \alpha \, |x|^p$ with $p~>~1$}. We now analyse the general stationary solution in Eq. (\ref{pss-gen}) for the case $V(x) = \alpha\, |x|^p$, with $p>1$. In this case $f(x) = -V'(x) = -\alpha\,p |x|^{p-1} {\rm sgn}(x)$. Since the general solution in Eq. (\ref{pss-gen}) is symmetric around $x=0$, i.e., $P(x) = P(-x)$, it is sufficient to consider $P(x)$ for $x \geq 0$.
Substituting $f(x) =  -\alpha\,p x^{p-1}$ with $x>0$ in Eq. (\ref{pss-gen}), we get
\begin{equation}\label{exact_P1}
P(x) = \frac{B_p}{b_p^2-x^{2p-2}} \exp{\left[- \frac{2\gamma}{\alpha\,p} \int_0^x \frac{y^{p-1}}{b_p^2-y^{2p-2}}\,dy \right]} \;,
\end{equation}
where the parameter $b_p$ is given by
\begin{eqnarray}\label{bp}
b_p = \frac{v_0}{\alpha \,p } \;
\end{eqnarray} 
and the overall normalisation constant $B_p$ is such that $\int P(x)\, dx = 1$. The solution is actually symmetric around $x=0$ with a finite support $[-b_p^{\frac{1}{p-1}},+b_p^{\frac{1}{p-1}}]$. We next analyse Eq. (\ref{exact_P1}) to see how the solution behaves near the upper support end at $x=x_+=b_p^{\frac{1}{p-1}}$. Clearly, by symmetry, it has the same behavior at the lower support $x_-=-b_p^{\frac{1}{p-1}}$. 
To proceed, we set $x = b_p^{\frac{1}{p-1}} - \epsilon$ where $\epsilon >0$ is small. The integral inside the exponential in Eq. (\ref{exact_P1}) then reads 
\begin{eqnarray}\label{def_I}
I(\epsilon) = \int_0^{b_p^{\frac{1}{p-1}}-\epsilon} \frac{y^{p-1}}{b_p^2 - y^{2p-2}}\,dy \;.
\end{eqnarray}
We next make a change of variable $y \to b^{\frac{1}{p-1}}-z$ in the integral in Eq. (\ref{def_I}), and expand the integrand to leading order for small $z$. Performing the integral gives, to leading order for small $\epsilon$, 
\begin{eqnarray}\label{I_asympt}
I(\epsilon) \approx -\frac{b_p^{\frac{2-p}{p-1}}}{2(p-1)}\, \ln \epsilon \;.
\end{eqnarray}
Substituting this result in Eq. (\ref{exact_P1}), we then find that the stationary solution behaves, near the upper edge, as
\begin{eqnarray}\label{asympt_pss}
P(x = b_p^{\frac{1}{p-1}} -\epsilon) \sim \epsilon^{\phi} \;, \;
\end{eqnarray}
with a non-universal (i.e., parameter dependent) exponent  
{
\begin{equation}\label{phi_p}
\phi = \frac{b_p^{\frac{2-p}{p-1}}}{\alpha\, p(p-1)}\, \gamma - 1 = \frac{1}{p(p-1)} \left( \frac{v_0}{p}\right)^{\frac{2-p}{p-1}} \frac{\gamma}{\alpha^{\frac{1}{p-1}}} - 1\;.
\end{equation}}
Clearly, for $p>1$, the exponent $\phi > - 1$, indicating that $P(x)$ in Eq. (\ref{asympt_pss}) is integrable near the edge. For $p=2$, setting $\alpha = \mu/2$, we recover $\phi = \gamma/\mu - 1$ as in Eq. (\ref{phi_exp}). For any $p>1$, clearly there is a shape transition in the $(p, \alpha)$ plane (see Fig.~\ref{Fig_Ph_Diag}) accross the line  
\begin{eqnarray}\label{gammac}
{\alpha=\alpha_c(p) =  \left(\frac{v_0}{p}\right)^{2-p}\left(\frac{\gamma}{p(p-1)} \right) ^{p-1}}
\end{eqnarray}
such that $\phi>0$ for $\alpha<\alpha_c$ and $-1<\phi<0$ for $\alpha>\alpha_c$. Thus, as in the $p=2$ case, there is a shape transition ({from converging edges to diverging edges}) for generic $p>1$ as $\alpha$ increases through $\alpha_c$. The existence of this critical line $\alpha_c \equiv \alpha_c(p)$ in the $(p,\alpha)$ plane, separating an active and a passive phase, is indeed the main new result of this section.

 \begin{figure}[t]
\centering
 \includegraphics[width = \linewidth]{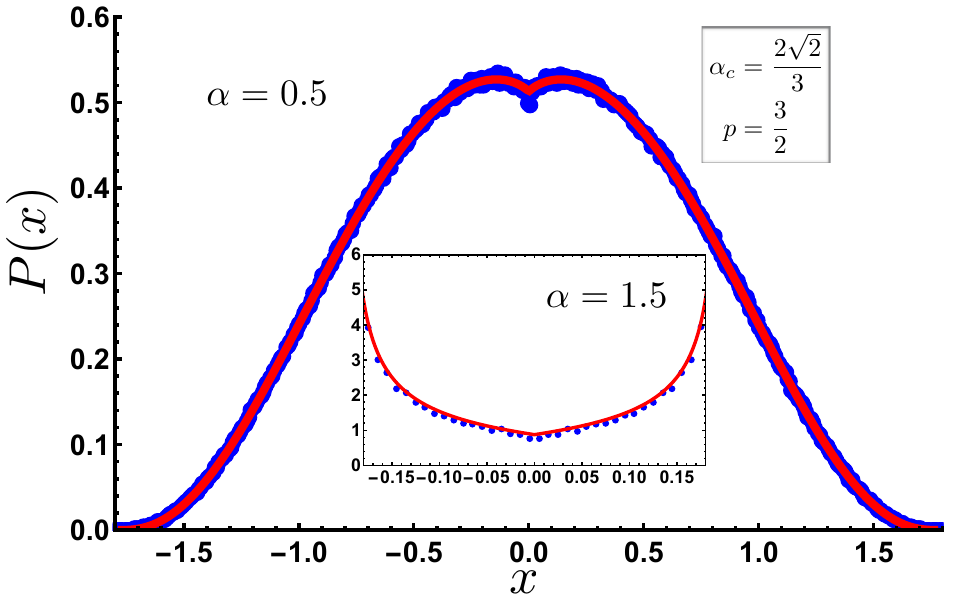} 
 \caption{{Numerical verification of the theoretical result of the steady state distribution in \eqref{exact_P1} for $V(x)=\alpha \,x^{3/2}$. We set the parameter values $\gamma = 1$ and $v_0 = 1$ and vary $\alpha$ to probe the shape-transition, as done for the harmonic case in Fig. \ref{fig:pss-harm}. In this case, the critical value of $\alpha_c$, from Eq. (\ref{gammac}) with $p=3/2$, is given by $\alpha_c =\frac{2\sqrt{2}}{3}=0.942809$. We considered two representative values of $\alpha = 0.5 < \alpha_c$ (main plot) and $\alpha = 1.5 > \alpha_c$ (inset) corresponding respectively to the passive and active-like phases. We find an excellent agreement between the numerical simulations (blue points) and the theoretical predictions (solid red lines) in Eq. (\ref{exact_P1}).}}
\label{fig6}
\end{figure}

Indeed the shape-transition for $p>1$, where the exponent $\phi$ in Eq. (\ref{phi_p}) changes sign, can be accessed by varying either
$\alpha$ or $\gamma$. Indeed from Eq. (\ref{phi_p}), the exponent $\phi$ only depends on the ratio {$\gamma/\alpha^{\frac{1}{p-1}}$. Thus, small (respectively large) $\gamma$ corresponds to large (respectively small) $\alpha$ for the same value of $\phi$}. While the phase diagram in Fig. \ref{Fig_Ph_Diag} is presented in the $(p,\alpha)$-plane, for $p>1$, we can equivalently probe this transition by varying $\gamma$, with $\alpha$ fixed. In fact, in the numerical simulations for $p=2$ (Fig. \ref{fig:pss-harm}) and $p=4$ (Fig. \ref{fig:pss-xpw4}), we indeed kept $\alpha$ fixed, while varying $\gamma$. In both cases, one clearly sees the transition and an excellent agreement between theoretical predictions and simulations.  

The behaviour near $x=0$ is also interesting. Performing an analysis around $x=0$, in the same was as was done near the right edge at $x = x_+ = b_p^{\frac{1}{p-1}}$, one finds for $p>1$ and small $\epsilon$
\begin{align}
P(\epsilon) \simeq \frac{B_p}{b_p^2} \left[1 - \frac{2 \gamma}{\alpha p^2 b_p^2}~|\epsilon|^p + \frac{1}{b_p^2}|\epsilon|^{2(p-1)} +O(|\epsilon|^{3p-2})\right]. \nonumber
\end{align}
{We see that, as $\epsilon \to 0$, the dominant sub-leading term is either the second (if $p>2$) or the third term (if $1 < p < 2$). In the former case, the sign of the sub-leading term is negative, while in the latter case it is positive. Consequently, the distribution $P(x)$, near $x=0$, is convex for $1<p < 2$ whereas for $p > 2$ it is concave, irrespective of whether $\alpha$ is greater or less than $\alpha_c(p)$. The behaviour for $1<p < 2$ is demonstrated in Fig.~\ref{fig6} where the distribution $P(x)$ is convex near $x=0$ for both $\alpha > \alpha_c$ and $\alpha <\alpha_c$.}

\vspace*{0.3cm}
\noindent{\it The marginal case $p=1$ where $V(x) = \alpha\,|x|$}. In this case, setting $f(x) = - \alpha~{\rm sign}(x)$ in Eq. (\ref{pss-gen}), we find that a   {non-localised stationary solution} exists only for $\alpha \leq v_0$ where it is given by
\begin{equation} \label{main_marginal}
P(x)=\frac{\gamma \alpha}{v_0^2-\alpha^2}\exp \left(- \frac{2\gamma \alpha}{v_0^2-\alpha^2} {|x|} \right),~-\infty \le x \le \infty \;.
\end{equation}  
Note that, unlike in the $p>1$ case, the stationary solution is no-longer supported on a finite interval but extends over the full space. From this solution in Eq. (\ref{main_marginal}), we also see that, when $\alpha$ approaches $v_0$ from below, the stationary distribution approaches to a delta function $P(x)=\delta(x)$ centred at the origin $x=0$. Indeed, for any $\alpha \geq v_0$, the system collapses to a delta function at $x=0$, at long times. This can be seen by analysing the Langevin equation (\ref{RTP}) which reads, in this case,  
\begin{eqnarray}\label{langevin_marginal}
\frac{dx}{dt} = -\alpha\, {\rm sgn}(x) + v_0\,\sigma(t) \;.
\end{eqnarray}
Imagine that the particle starts at some position $x_0>0$ with positive velocity $\sigma(0) = +1$. If this velocity stays at $+1$, and $\alpha > v_0$, the particle gets driven towards the origin with $x(t) = x_0-(\alpha-v_0)\,t$. Thus, in a finite time, it will arrive at the origin and subsequently it can not escape the origin and eventually collapses to the origin, giving rise to a delta function at $x=0$. If the initial velocity is negative $\sigma(0)=-1$, it arrives at the origin even faster and, consequently, the same collapse phenomenon to the origin occurs.

\begin{figure}[t]
\includegraphics[width=\linewidth]{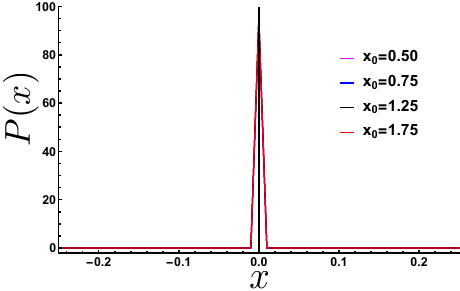}
\caption{Numerical verification of the theoretical prediction for the stationary distribution $P(x) = \delta(x)$ for $V(x) = \alpha \,|x|^p$, with $p=1/2$, $\alpha = 2$, $\gamma = 1$ and $v_0=1$, for different choices of the initial position $x_0$.}\label{Fig_pleq1}
\end{figure}

\vspace*{0.3cm}
\noindent{\it The case $p<1$.} This case is very similar to the marginal case $p=1$ with $\alpha>v_0$, as discussed above. Indeed, in the large time limit, the stationary solution is given by a delta-function at the origin $P(x) = \delta(x)$, for all $p<1$ and arbitrary $\alpha>0$. 
To see this, we again consider the Langevin equation that now reads
\begin{eqnarray}\label{langevin_pleq1}
\frac{dx}{dt} = -\alpha~p \,{\rm sgn}(x) |x|^{p-1} + v_0\,\sigma(t) \;.
\end{eqnarray}
We now consider a similar argument as in the marginal case. We first assume that the particle starts at $0<x_0<x_+=b_p^{1/(p-1)}$ with $b_p = v_0/(\alpha p)$. In this case, irrespective of the sign of $\sigma(t)$, the particle feels a force toward the origin, which gets stronger and stronger as $x \to 0$. Thus, in this case, the particle eventually collapses to $x=0$. If the initial position $x_0>x_+$ and if the initial noise is positive the particle initially feels a force away from the origin. However, when the noise changes sign, it reverses the direction. Since the walk, at long times, is expected to be recurrent, at some time, it will definitely arrive at $x=x_+$, and with probability one it will cross to the other side. Once $x(t)<x_+$, then again the particle gets strongly attracted to the origin, leading to an eventual collapse to the origin. A similar argument can be made for $x_0<0$. 
Thus, in all cases, we would expect that, for $p<1$, no matter what the starting position is, the stationary solution is just $P(x) = \delta(x)$. We have verified this numerically for different values of $p<1$ and different values of $x_0$. In Fig. \ref{Fig_pleq1} we show the stationary $P(x)$ for $p=1/2$ and for different values of $x_0$ (setting $\alpha = 2$, $v_0=1$, in which case $x_+ = 1$ since $p=1/2$).        

Thus all the cases $p>1$, $p<1$ and $p=1$ can be summarised succinctly in the phase diagram in the $(p,\alpha)$ plane as shown in Fig. \ref{Fig_Ph_Diag}.

\section{Relaxation to the steady state}
\label{relax-harmonic}

After having discussed the stationary state in the previous section, it is natural to ask next how the time dependent probability distribution relaxes to this stationary state at long times. In principle, this relaxation dynamics can be studied for arbitrary confining potentials $V(x)$. However, for simplicity we focus here on the simple harmonic case with $V(x) = \mu x^2/2$, i.e.,  $f(x)=-\mu x$ for which Eqs. \eqref{eqn:P1} and \eqref{eqn:P2} can be solved explicitly, as we see later. 

\begin{figure}
\includegraphics[width = 0.8\linewidth]{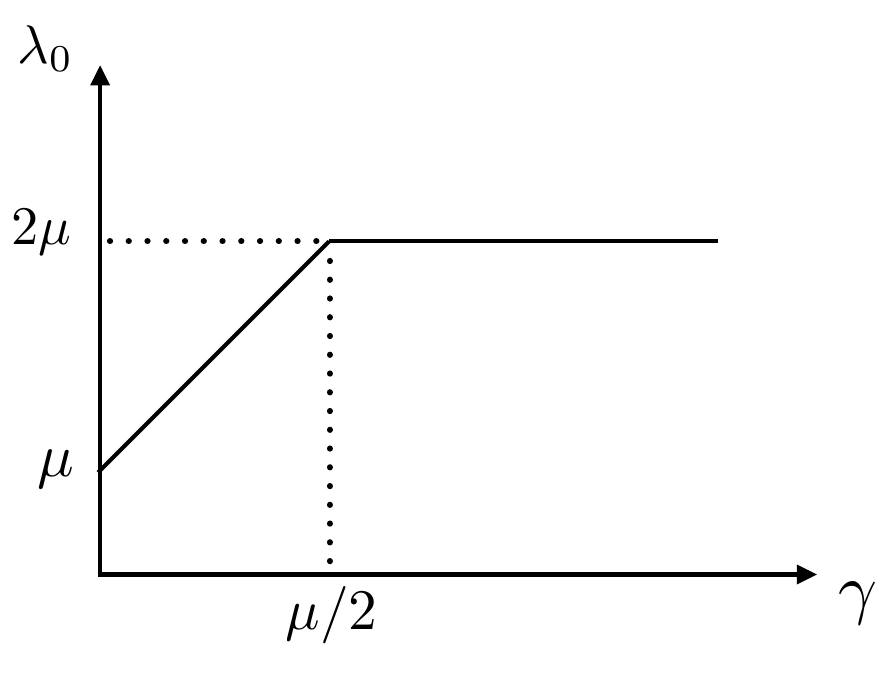}
\caption{Plot of $\lambda_0$ as a function of $\gamma$ for $x_0 = 0$ [see Eq. (\ref{res_lambda0})].} \label{Fig_lambda0}
\end{figure}

\begin{figure}[h!]
\centering
\includegraphics[scale=0.85]{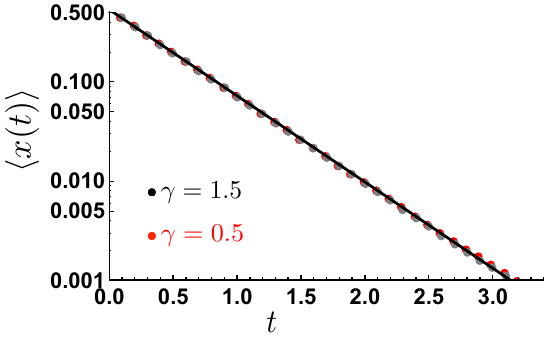} \includegraphics[scale=0.8]{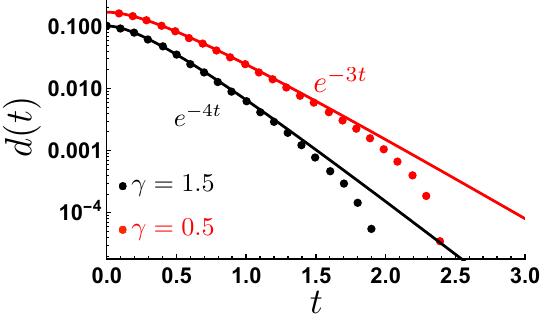} 
\caption{Relaxation in the harmonic potential $V(x) = \mu/2\,x^2$ with choices of parameters $\mu=2.0$ and $v_0=1$, and for two different values
 of $\gamma = 0.5$ (red) and $\gamma = 1.5$ (black). The left panel shows the decay of $\langle x(t)\rangle$ as a function of time, starting from $x_0 = 0.53$. Clearly, the relaxation is a pure exponential, as predicted in Eq.~(\ref{av-x_t}), independently of $\gamma$. In the right panel, we plot $d(t) = {\rm Var}(\infty) - {\rm Var}(t)$ where ${\rm Var}(t) = \langle x(t)^2 \rangle-\langle x(t) \rangle^2$ is the variance at time $t$. According to the theoretical prediction in Eqs.~(\ref{d_of_t}) and (\ref{lambda}), $d(t) \sim e^{- \lambda t}$ with $\lambda = \min(\mu + 2 \gamma, 2 \mu)$. Since $\mu = 2.0$, we expect a change of behaviour in the decay exponent $\lambda$ at $\gamma = 1$: $\lambda = 2 + 2 \gamma$ if $\gamma<1$ while $\lambda = 4$ if $\gamma > 1$. The simulations for $\gamma = 0.5 < 1$ and $\gamma = 1.5 > 1$ are in excellent agreement with this theoretical prediction.}
\label{fig:relaxation}
\end{figure}

{
Before computing the full time dependent solution $P(x,t)$, it is useful and simpler to compute the time evolution of the moments of the distribution. This already provides a clue as to how the system relaxes to the stationary state. In particular it determines already the leading relaxation time scale. In fact, just the first two moments, e.g., the mean and the variance, are enough for this purpose. 
The first two moments of $x(t)$ can be computed directly from the Langevin equation $dx/dt = -\mu x + v_0 \sigma(t)$ where $\langle \sigma(t) \rangle = 0$ and $\langle \sigma(t) \sigma(t') \rangle = \exp(-2 \gamma |t-t'|)$. Integrating this Langevin equation, one gets
\begin{eqnarray}\label{sol_Langevin}
x(t) = x_0 \, e^{-\mu\,t} + v_0 \, e^{-\mu t}\int_0^t dt_1 \sigma(t_1)\,e^{\mu t_1} \;.
\end{eqnarray}
Computing the first two moments gives
\begin{align}
\langle x(t) \rangle&= x_0~e^{-\mu t}, \label{av-x_t} \\
\langle x(t)^2 \rangle-\langle x(t) \rangle^2 &=v_0^2 \left[\frac{1}{2 \gamma  \mu +\mu ^2}+\frac{2 e^{-t (\mu+2 \gamma)}}{4 \gamma ^2-\mu ^2} \right. \nonumber \\ 
&~~~~~~~~~~~~~~~~~~~~~~~~\left.+\frac{e^{-2 \mu  t}}{\mu  (\mu -2
   \gamma )}\right], 
   \label{var-x_t0}
\end{align}
where $\mu\not=2\gamma$ in the second equation. For $\mu=2\gamma$, we get
\begin{eqnarray}\label{mu_2g0}
\langle x(t)^2 \rangle-\langle x(t) \rangle^2 = \frac{v_0^2}{8\gamma^2}\left[1 - e^{-4 \gamma t} - 4 \gamma\,t \,e^{-4 \gamma t} \right] .
\end{eqnarray}
Thus the variance, unlike the mean is independent of the initial position $x_0$. 

For generic $x_0 \neq 0$ the first moment decays as $\sim e^{-\lambda_0\,t}$, with $\lambda_0=\mu$,  indicating that the largest relaxation time is $\mu^{-1}$. For $x_0=0$, the average $\langle x(t) \rangle$ remains zero at {all}  times, by symmetry. Hence the lowest non-zero cumulant is the variance ${\rm Var}(t) = \langle x(t)^2 \rangle-\langle x(t) \rangle^2$. 
We see from Eqs.~(\ref{var-x_t0}) and \eqref{mu_2g0} that the variance decays to its stationary value ${\rm Var}(\infty)$ at large times as 
\begin{eqnarray}\label{d_of_t}
d(t) = {\rm Var}(\infty) - {\rm Var}(t) \sim {e^{-\lambda_0\,t}} \;, 
\end{eqnarray}
{where the slowest relaxation mode is given by 
\begin{eqnarray}\label{lambda}
\lambda_0 = \min(\mu + 2 \gamma, 2 \mu) \;.
\end{eqnarray}}
Hence, 
to summarise (see Fig. \ref{Fig_lambda0}),  
\begin{align}\label{res_lambda0}
\lambda_0=
\begin{cases}
\mu & \text{if,}~x_0 \neq 0 \\
\min[\mu+2 \gamma, 2 \mu]  & \text{if,}~x_0 = 0 \;.
\end{cases}
\end{align}

Thus for $x_0=0$, there is a transition in the relaxation behaviour as one increases the jump rate $\gamma$ for fixed $\mu$. At small $\gamma$ the relaxation is given by $\mu+2\gamma$. When $\gamma$ becomes larger than $\mu/2$, the relaxation is given by $2 \mu$ [see Fig.~\ref{Fig_lambda0} and Fig.~\ref{fig:relaxation} for a numerical verification of this change of behavior at $\gamma=\gamma_c=\mu/2$]. 
Note that for a passive Brownian particle (i.e.,  $\gamma \to \infty$ and $v_0 \to \infty$ limit keeping $v_0^2/\gamma =2D$ fixed)  in a harmonic potential (Ornstein-Uhlenbeck case), the relaxation behaviour is $\sim e^{-\mu \,t}$ for a generic $x_0 \neq 0$, and $\sim e^{-2\mu\,t}$ for $x_0=0$, consistent with the $\gamma \to \infty$ limit of Eq.~(\ref{res_lambda0}).

}

\begin{figure}
\includegraphics[width=3.4in]{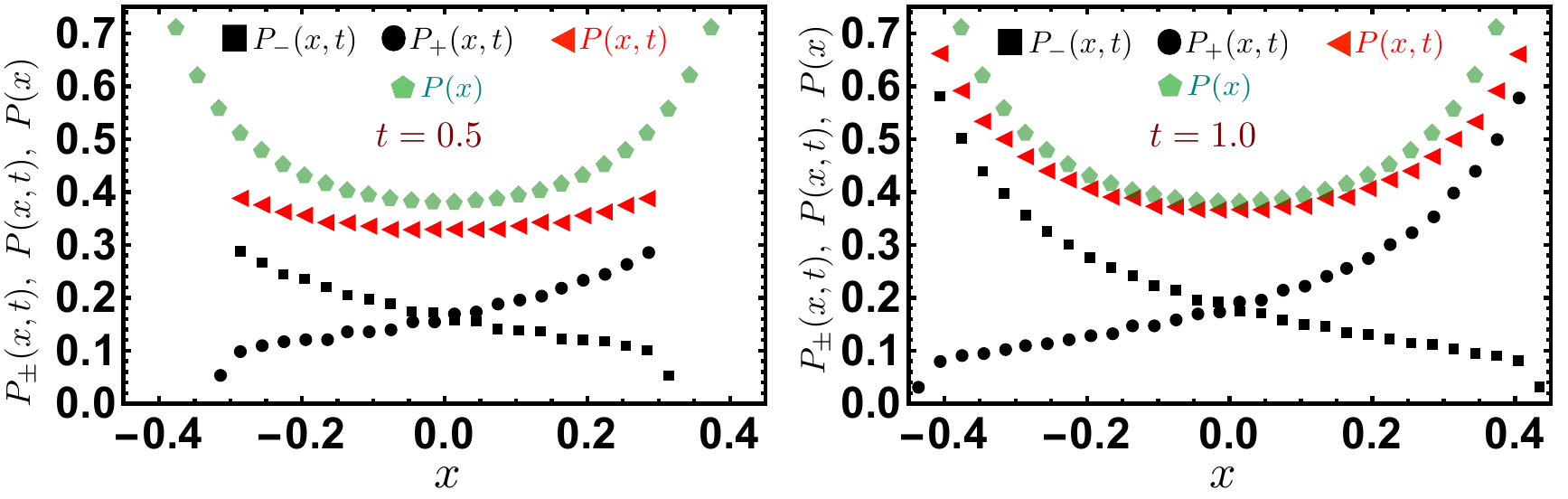}
\caption{{Plots of the probability distributions $P_\pm(x,t)$ and $P(x,t)$ at times $t=0.5$ and $t=1.0$. The green filled polygons represent the steady state obtained at large times. At finite time there are delta function peaks at the edges of $P(x,t)$ \emph{i.e.} at $x_\pm(t)=\pm (v_0/\mu) [1-\exp(-\mu t)]$ (not shown in the figure) whose strength decreases to zero as $e^{-\gamma t}$. Other parameters of the plot are: $x_0=0,~\gamma=0.5,~\mu=2.0$ and $v_0=1.0$.}}
%
\label{relaxation-plot0}
\end{figure}

{ 
Having computed the temporal behavior of the first two moments, we now consider the full time dependent probability distribution $P_\pm(x,t)$. 
 From the discussion above, we have seen that the initial condition $x_0=0$ is special and in some sense more interesting, since the relaxation time scale $\lambda_0$ undergoes a transition as a function of $\gamma$. Hence, in the following, we focus on the solution of the pair of Fokker-Planck equations   \eqref{eqn:P1} and \eqref{eqn:P2} with   $f(x)=-\mu x$ and the following boundary and 
 initial conditions
\begin{align}
&P_\pm(|x| \to \infty, t)=0,~~~\text{and}~~~P_\pm(x,t=0)=\frac{1}{2}\delta(x) \;.
\label{eqn:P2-harmonic-bc}
\end{align}

} 
Note that this choice of initial condition ensures that the particle, initially at the origin at $t=0$, always remains confined in the finite box $[-v_0/\mu,v_0/\mu]$. This can be easily seen by solving 
\eqref{RTP} for the {extreme cases,}  $\sigma(t) = 1$ or $-1$ for $0 \le t \le \infty$,  {which yields the deterministic solution $x(t)=\pm (v_0/\mu) [1-\exp(-\mu t)]$}. So the distributions $P_\pm(x,t)$ have a finite support $x \in [-v_0/\mu,~v_0/\mu]$ for all time $t$. {The total probability density at time $t$ is given by $P(x,t)=P_+(x,t) + P_-(x,t)$. 
To see how $P_\pm(x,t)$ and their sum $P(x,t)$ approach their respective steady state forms given in Eqs. \eqref{P_pm-harm} and \eqref{harm-ss}, we first provide the results from simulations in Fig.~\ref{relaxation-plot0}. In this figure, $P_\pm(x,t)$ are plotted as a function of $x$ at two different times $t=0.5$ and $t=1.0$ (shown by black circles and black squares), with parameters $\gamma=1/2$, $\mu=2$, and $v_0=2$. Hence, the distribution has the support over $[-1/2,1/2]$. Also, their sum $P(x,t)$ is plotted by red triangles and the green polygons denote the steady state $P(x)$. We see clearly from this figure that as time increases,  $P(x,t)$ approaches the stationary $P(x)$. Below, we compute analytically this relaxation of $P_\pm(x,t)$ to the steady state at late times.

}

To solve the time-dependent equations Eqs.~\eqref{eqn:P1} and \eqref{eqn:P2}, with $f(x) = - \mu x$, it is convenient to first take
their Laplace transforms with respect to time. We thus define
\begin{eqnarray}
\tilde P_{\pm}(x,s) = \int_0^\infty e^{-st} \, P_{\pm}(x,t) \, dt \label{def_Laplace} \;.
\end{eqnarray}
Taking Laplace transforms of Eqs.~\eqref{eqn:P1} and \eqref{eqn:P2}, integrating by parts the left hand side using the initial conditions in Eq. (\ref{eqn:P2-harmonic-bc}), we get 
\begin{align}
\begin{split}
& (\mu x-v_0)\frac{\partial \tilde{P}_{+}}{\partial x} +(\mu-\gamma -s) \tilde{P}_{+} +  \gamma \tilde{P}_{-} = -  \frac{1}{2}\delta(x) \;,  \\
&(\mu x+ v_0)\frac{\partial \tilde{P}_{-}}{\partial x}+(\mu-\gamma -s) \tilde{P}_{-} +  \gamma \tilde{P}_{+} = -  \frac{1}{2}\delta(x) \;,   
\end{split}
\label{eqn:P2-harmonic-laplace} 
\end{align}
for $x \in [-v_0/\mu,~v_0/\mu]$. It is convenient to first make a change of variable
\begin{equation}
x=\frac{v_0}{\mu}(1-2z)\,, \label{transf}
\end{equation}
that transforms the domain $x\in[-v_0/\mu,v_0/\mu]$ to $z\in[0,1]$. 
We get, 
\begin{align}
\begin{split}
& \left[ z \frac{\partial }{\partial z} +(1-\bar{\gamma}-\bar{s}) \right]\tilde{P}_{+} = -  \bar{\gamma} \tilde{P}_{-} -  {\frac{1}{4v_0}} \delta \left(z-\frac{1}{2}\right),  \\
&\left[ (z-1) \frac{\partial }{\partial z} +(1-\bar{\gamma}-\bar{s}) \right]\tilde{P}_{-}  =-  \bar{\gamma} \tilde{P}_{+} -{\frac{1}{4 v_0}}\delta \left(z-\frac{1}{2}\right),   
\end{split}
\label{eqn:P2-harmonic-laplace-z} 
\end{align}
where $\tilde{P}_\pm(z,s)\equiv \tilde{P}_\pm(v_0(1-2z)/\mu,~s)$ and 
\begin{align}
\bar{\gamma}= \gamma/ \mu,~~\text{and}~~\bar{s}=s/\mu \;. \label{gamma-bar} 
\end{align}
{Let us remark that the original normalisation condition 
\begin{eqnarray} \label{normalisation_1}
\int_{-v_0/\mu}^{v_0/\mu} [P_+(x,t) + P_-(x,t)]\, dx = 1
\end{eqnarray}
translates, in terms of the $z$ coordinate defined in Eq.~(\ref{transf})
\begin{eqnarray}\label{normalisation_2}
\int_0^1[P_+(z,t) + P_-(z,t)]\, dz = \frac{\mu}{2 v_0} \;.
\end{eqnarray}
Taking the Laplace transform with respect to $t$ yields
\begin{eqnarray}\label{normalisation_3}
\int_0^1 [\tilde P_+(z,s) + \tilde P_-(z,s)]\, dz = \frac{\mu}{2 v_0 s} = \frac{1}{2v_0 \bar s} \;,
\end{eqnarray}
with $\bar s = s/\mu$. We further choose, for convenience and without any loss of generality,
\[ v_0=1 \;.\]} 
We now solve these equations separately for $z < 1/2$ and $z > 1/2$. Due to presence of the delta-function on the right hand side of Eq. (\ref{eqn:P2-harmonic-laplace-z}), the solutions in these two disjoint regions are related via the matching condition at $z=1/2$. 
\begin{align}
\begin{split}
\tilde{P}_+(z=1/2 + \epsilon, \bar{s}) - \tilde{P}_+(z=1/2-\epsilon, \bar{s}) &= {-1/2} \;,  \\ 
\tilde{P}_-(z=1/2 + \epsilon, \bar{s}) - \tilde{P}_-(z=1/2-\epsilon, \bar{s}) &= {1/2} \;,
\end{split}
\label{discont-cond}
\end{align}
where $\epsilon \to 0^+$.
{ 
Note that these jump  discontinuities at $z=1/2$ in the Laplace transforms ${\tilde P}_{\pm}(z,s)$ originates from the delta function at $x=0$  in the initial condition
of $P_{\pm}(x,0)$ in   Eq. (\ref{eqn:P2-harmonic-bc}).  However, these jumps  at $z=1/2$  or $x=0$ in the Laplace transforms do not mean that the
$P_{\pm}( x,t)$ in real time also have discontinuities at $x=0$.  Indeed, at all times $t$, the function $P_{\pm}(x,t)$ are both
continuous at $x=0$, as seen, e.g., in Fig. \ref{relaxation-plot0}.
In addition,} from Eqs. \eqref{eqn:P2-harmonic-laplace-z}, it is clear that the solution
has the following symmetry property 
\begin{equation}
\tilde{P}_\pm(z, \bar{s}) = \tilde{P}_\mp(1-z, \bar{s}) \;, \label{symmetry}
\end{equation} 
and hence it is sufficient to solve Eqs. (\ref{eqn:P2-harmonic-laplace-z}) in one region only, say $0\leq z \leq 1/2$. Using this symmetry (\ref{symmetry}), we can further replace $\tilde{P}_+(z=1/2+\epsilon,\bar{s})$ by $\tilde{P}_-(z=1/2-\epsilon, \bar{s})$ in the first line of Eq. (\ref{discont-cond}), which then reads
\begin{equation}
\tilde{P}_-(z=1/2 - \epsilon, \bar{s}) - \tilde{P}_+(z=1/2-\epsilon, \bar{s}) = {-1/2}. \label{discont-cond-single}
\end{equation}

We now focus on the $z<1/2$ case with the boundary condition as in Eq. (\ref{discont-cond-single}). From the two first order differential equations in \eqref{eqn:P2-harmonic-laplace-z} one can eliminate 
$\tilde{P}_-$ (respectively $\tilde{P}_+$) to get a second order differential equation for $\tilde{P}_+$ (respectively $\tilde{P}_-$)
\begin{align}
&z(1-z) \frac{\partial^2 \tilde{P}_\pm}{\partial z^2} +[c_\pm - (1+a+b)z ]\frac{\partial \tilde{P}_\pm}{\partial z} - a \,b \tilde{P}_\pm = 0, \label{p-pm-2nd-order-diffeq}  \\
&\text{where,} \nonumber \\
& a=1-\bar{s},~~b=1-2\bar{\gamma}-\bar{s}, \\
&c_+=2-\bar{\gamma}-\bar{s}, ~~\text{and}~~
c_-=1-\bar{\gamma}-\bar{s} \;. \label{constants-abc}
\end{align}
{
Equation \eqref{p-pm-2nd-order-diffeq} is a standard hypergeometric differential equation which has two linearly independent solutions. We show in the appendix~\ref{appendixA} that the solution to this differential equation can be uniquely determined by using the normalization condition in Eq.~\eqref{normalisation_3} (setting $v_0=1$) and the jump conditions in Eq.~\eqref{discont-cond}.   Skipping details, the final solution for $\tilde{P}(z,s)=\tilde{P}_+(z,s) + \tilde{P}_-(z,s)$  reads
\begin{equation}
\tilde{P} (z,s) = {B}(s)\, z^{\bar\gamma +\bar{s} -1} F(1-\bar\gamma, \bar\gamma, \bar\gamma+\bar{s}, z), 
\label{Pzs.0}
\end{equation}
where  {$F(a,b,c,z) \equiv \, _2\,\!F_1(a,b;c; z)$ }is a standard hypergeometric function.  The amplitude $B(s)$ is given by
\begin{equation}
{B}(s)=2^{ 2(\bar\gamma +\bar{s})  -3} 
\frac{\Gamma(\bar{s}/2)\Gamma(\bar\gamma + (1+\bar{s})/2)}{\sqrt{\pi}\Gamma(\bar\gamma+\bar{s})}, 
\label{A-s}
\end{equation}
and we recall $\bar{s}=s/\mu$ and $\bar\gamma=\gamma/\mu$. 

As a first check of this result, we note that 
in the $s \to 0$ limit we get
\begin{equation}
\tilde{P}(z,s) \to \frac{1}{s}~\frac{\mu}{2}P(z),~~\text{where}~~P(z)=\frac{[z(1-z)]^{\bar{\gamma}-1}}{B(\bar{\gamma},\bar{\gamma})} \;,
\end{equation}
with $B(\alpha,\beta)$ being the beta function.
Expressing $z$ in terms of $x$ using Eq. (\ref{transf}), and inverting the Laplace transform trivially, we then recover the steady state distribution given in Eq.~\eqref{harm-ss}. 

In the opposite limit $s \to \infty$,  the Eq.~\eqref{Pzs.0} along with Eq.~\eqref{A-s}  reduces to $\tilde{P}(z,s) \to (2 z)^{\bar{\gamma}+\bar{s}-1}/2$ where we recall that $0\leq z \leq 1/2$. By inverting the Laplace transform and using the transformation in Eq.~\eqref{transf}, one finds that  at small times $P(x,t)  \sim  \frac{e^{-\gamma t}}{2} \delta \left(x-\bar{x}(t) \right)$ for $x>0$ where 
\[\bar{x}(t)=\frac{1}{\mu}(1-e^{\mu t}).\]
A similar analysis for $1/2 \leq z \leq 1$, shows that $P(x,t) $ has a delta function at position $-\bar{x}(t)$ of strength $e^{-\gamma t}$ for $x<0$. Hence, at small times the distribution is  
\[
P(x,t)  \sim \frac{e^{-\gamma t}}{2}\left[ \delta \left(x-\bar{x}(t) \right) +\delta \left(x+\bar{x}(t) \right)\right].
\]

To derive the relaxation to this steady state, one needs to investigate the poles of $B(s)$ in Eq.~\eqref{A-s}
 in the complex-$s$ plane. The poles in $B(s)$ are either the poles of $\Gamma(\bar{s}/2)$ or that of $\Gamma(\bar\gamma+(1+\bar{s})/2)$. This gives rise to two families of poles: the first one occurs when $\bar{s}=\frac{s}{\mu}=-2n$ for $n=0,1,2, \ldots$, while the second one occurs when $\bar{s}=\frac{s}{\mu}=-(1+2\bar{\gamma}+2m)$ for $m=0,1,2, \ldots$.   The solution corresponding to $n=0$, i.e. a pole at $s=0$, corresponds to the steady-state solution, while the slowest relaxation mode to the steady-state is decided by the zero closest to $s=0$ on the negative $s$-axis. This can be either $s = - 2 \mu$ (corresponding to $n=1$ in the first family) or to $s= -(\mu + 2 \gamma)$ (corresponding to $m=0$ in the second family). Thus, comparing these two, we see that the slowest relaxation mode corresponds to 
\begin{equation}
\label{slowest_mode}
s=- \min(\mu+2\gamma, 2\mu) ,
\end{equation}
and hence, we get
\begin{equation}\label{exp_1}
P(z,t)-P(z) \propto e^{-\lambda_0 t},~~\text{for}~t \to \infty,
\end{equation}
with $\lambda_0 = \min[\mu + 2 \gamma, 2 \mu]$, as given in the second line of Eq. (\ref{res_lambda0}), corresponding to $x_0 = 0$. 
This result is, therefore,  completely consistent with the results from the moments. 
Of course, it is straightforward also to evaluate the residues at these poles, and obtain a formal infinite series of $P(x,t)$ providing the full time dependent solution. While this is straightforward, it is a bit cumbersome, and hence, we do not give the explicit expression here.

}

\section{First-passage properties}
\label{FP}
In this section, we investigate the first-passage properties of the active particle in a confining potential. The first-passage 
probability density $F_{\rm fp}(t,x_0)$ is defined as follows: for a particle starting at $x_0 > 0$, $F_{\rm fp}(t,x_0)\, dt$ denotes the
probability that the particle crosses the origin $x=0$ for the first time in the interval $[t, t+dt]$~\cite{satya_review, redner_book, bray_review}. It is conveniently computed from the time-derivative 
$F_{\rm fp}(t,x_0) = -\partial_t Q(x_0,t)$ where $Q(x_0,t)$ is the survival probability of the particle up to time $t$, starting at $x_0$, in presence of an absorbing wall at the origin $x=0$. Let $Q_\pm(x_0,t)$ denote the survival probability of the particle starting at $x_0$ with, respectively, positive or negative velocity. Assuming that, initially, positive and negative velocities occur with equal probability $1/2$, we have $Q(x_0,t) = (Q_+(x_0,t) + Q_-(x_0,t))/2$. 
To analyse these survival probabilities, it is well known that, studying their backward Fokker-Planck equations is more convenient than the corresponding forward Fokker-Planck equations \cite{bray_review}. In this backward Fokker-Planck approach, one treats the initial position $x_0$ as a variable and by analysing the stochastic moves in the {\it first} time step using the Langevin equation $dx/dt = f(x) + v_0 \,\sigma(t)$, one can easily write down the backward Fokker-Planck equations as  
\begin{align}
\partial_t Q_{+}(x_0,t) =& - \gamma Q_{+}(x_0,t) + \gamma Q_{-}(x_0,t) \nonumber \\ 
&+  [f(x_0)+v_0] ~\partial_{x_0} Q_{+}(x_0,t) ,  \label{eqnQ+} \\
\partial_t Q_{-}(x_0,t) =&  \gamma Q_{+}(x_0,t) - \gamma  Q_{-}(x_0,t)   \nonumber \\ 
&+  [f(x_0)-v_0] ~\partial_{x_0} Q_{-}(x_0,t) \;. \label{eqnQ-} 
\end{align}
These equations have to be solved in the regime $x_0 \in [0,+\infty)$ with the boundary conditions: a) $Q_-(x_0=0,t)=0$. This is because if the particle starts at the origin with a negative velocity it can not survive up to any finite time $t$. However, if it starts at $x_0=0$ with a positive velocity, it can survive up to any finite time $t$. Hence $Q_+(x_0=0,t)$ is not specified. b) $Q_\pm(x_0 \to \infty,t)=1$ which follows from the fact that, if the particle starts initially at $x_0 = + \infty$, then irrespective of its initial velocity, it will survive with probability 1 up to any finite time $t$. Furthermore, we specify the initial conditions $Q_{\pm}(x_0,t=0) = 1$.

To solve Eqs. (\ref{eqnQ+}) and (\ref{eqnQ-}), it is convenient to take the Laplace transform with respect to $t$, $\tilde{Q}_\pm(x_0,s) = \int_0^\infty e^{-st}Q_\pm(x_0,t)$. Using the initial condition $Q_\pm(x_0,t=0)=~1$ we find 
 \begin{align}
\label{eqnQtilde}
-1+s\tilde{Q}_+(x_0,s)  =& - \gamma \tilde{Q}_+(x_0,s)+ \gamma \tilde{Q}_-(x_0,s) \nonumber \\  
&+  [f(x_0)+v_0] ~\partial_{x_0} \tilde{Q}_+(x_0,s) ,\\
-1+s\tilde{Q}_-(x_0,s) =&  \gamma \tilde{Q}_+(x_0,s) - \gamma  \tilde{Q}_-(x_0,s)  \nonumber \\  
&+  [f(x_0)-v_0] ~\partial_{x_0} \tilde{Q}_-(x_0,s)\;.
\end{align}
The boundary conditions a) and b) translate respectively to $\tilde{Q}_-(x_0=0,s)=0$ and $\tilde{Q}_\pm(x_0 \to \infty,s)=1/s$. 
Making a shift 
\begin{eqnarray}
\label{shift}
\tilde{Q}_\pm(x_0,s)=1/s + \tilde{q}_\pm(x_0,s) 
\end{eqnarray}
these inhomogeneous equations (\ref{eqnQtilde}) can be transformed 
to a homogeneous form 
\begin{align}
\begin{split}
&[f(x_0)+v_0] ~\partial_{x_0} \tilde{q}_+(x_0,s) r \\ 
&~~~~~~- (\gamma+s)~ \tilde{q}_+(x_0,s)+ \gamma~\tilde{q}_-(x_0,s) =0  ,\\
&[f(x_0)-v_0] ~\partial_{x_0} \tilde{q}_-(x_0,s)   \\ 
&~~~~~~+  \gamma ~\tilde{q}_+(x_0,s) - (\gamma+s)~ \tilde{q}_-(x_0,s) =0,
\end{split}
\label{eqnqtilde}
\end{align}
with the boundary conditions $\tilde{q}_-(x_0=0,s)=-1/s$ and $\tilde{q}_\pm(x_0 \to \infty,s)=0$. 

While these equations hold for a general force $f(x) = -V'(x)$, they are hard to solve for arbitrary potentials. However, for the harmonic case $V(x) = (\mu/2) \, x^2$, they can be explicitly solved as we show below. Substituting $f(x) = -\mu \,x$ in these equations, they can be re-written in the following convenient form
\begin{figure}[t]
\centering
 \includegraphics[width=\linewidth]{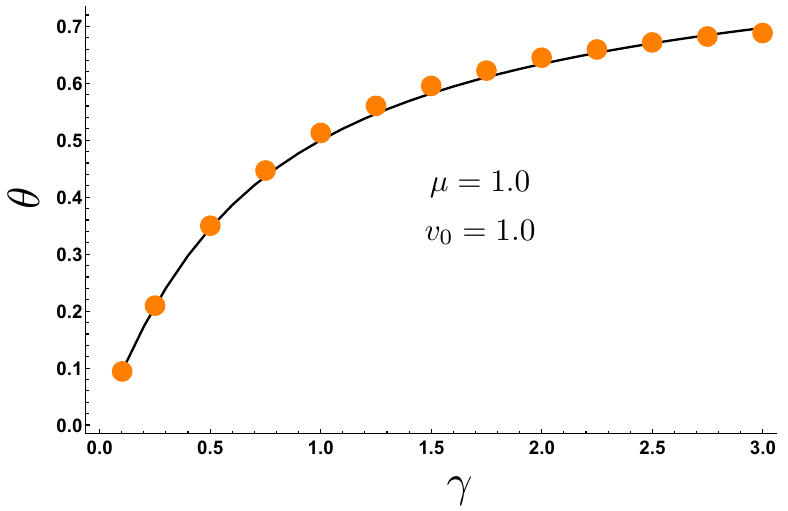} 
 \caption{We measure the decay exponent $\theta$,  characterizing the exponential decay of the survival probability
 $Q(x_0,t) \sim e^{-\theta\,t}$ for the harmonic potential $V(x) = \mu x^2/2$ with parameters $\mu = 1$, $v_0 = 1$, as a function 
 of $\gamma$. The solid line corresponds to the root of the hypergeometric function in Eq. (\ref{root1}) while the filled red dots
 correspond to the numerically obtained values, showing excellent agreement.}
\label{fig:theta-harm}
\end{figure}
\begin{align}
\label{eqnqtilde-harm-1}
&\hat L_+~ \tilde{q}_+(x_0,s)=- \gamma~\tilde{q}_-(x_0,s)   ,\\
&\hat L_-~\tilde{q}_-(x_0,s)  =- \gamma~\tilde{q}_+(x_0,s) ,
\label{eqnqtilde-harm-2} 
\end{align}
where the pair of operators $\hat L_\pm$ are given by 
\begin{align} \label{Lpm}
&\hat L_\pm=[-\mu x_0\pm v_0] ~\partial_{x_0}- (\gamma+s)
\end{align}
and the boundary conditions are as before, 
\begin{align}
&\tilde{q}_-(x_0=0,s)=-1/s \label{bc1} \\
&\tilde{q}_\pm(x_0 \to \infty,s)=0 \label{bc2} \;.
\end{align}
Applying $\hat L_-$ on both sides of Eq.~\eqref{eqnqtilde-harm-1} and using Eq. (\ref{eqnqtilde-harm-2}), one can eliminate $\tilde q_-(x_0,s)$ and 
obtain a closed equation for $\tilde q_+(x_0,s)$ only. Similarly, applying $\hat L_+$ on both sides of Eq. (\ref{eqnqtilde-harm-2}) and using Eq.~\eqref{eqnqtilde-harm-1} one obtains a closed equation for $\tilde q_-(x_0,s)$ only. These equations read, for $x_0 \in [0, +\infty)$
\begin{eqnarray}
(\mu^2x_0^2 - v_0^2) \partial_{x_0}^2\tilde{q}_\pm&+&[\mu(\mu+2\gamma+2s)x_0\pm\mu v_0] \partial_{x_0}\tilde{q}_\pm \nonumber \\ 
&+&(s^2+2 s \gamma) \tilde{q}_\pm=0 \;. \label{q_pm-tilde-harm}
\end{eqnarray}
We now make the customary change of variable $\mu x_0 = v_0(1-2z)$ (we will see that this change of variable transforms the differential equations in standard hypergeometric differential equations). The domain $0 \le x_0 < \infty$ transforms to $-\infty <z\le\frac{1}{2}$ for $z$. The equations in the $z$ variable read
\begin{equation}
z(1-z) \partial_z^2\tilde{q}_\pm + [c'_\pm - (a'+b'+1)z] \partial_z\tilde{q}_\pm - a' b' \tilde{q}_\pm=0,~~\label{qtilde-hyper-G}\\
\end{equation}
where
\begin{eqnarray}
&&a'=\frac{s}{\mu},~~b'=\frac{s+2\gamma}{\mu},~~\text{and} \label{a-b-pm}\\
&&c'_+=1+\frac{\gamma +s }{\mu} \;, \quad c'_-=\frac{\gamma +s }{\mu} \;.\label{abc}
\end{eqnarray}
{Note that the parameter $v_0$ is completely eliminated in Eq.~\eqref{qtilde-hyper-G}.}
It is sufficient to solve the equations (\ref{qtilde-hyper-G}) for one of them, either $\tilde q_+$ or $\tilde q_-$. It turns out to be more convenient to solve for $\tilde q_-$, for which the boundary conditions, using (\ref{bc1}) and (\ref{bc2}), are simply 
\begin{eqnarray}
&&\tilde q_-(z=1/2,s) = -\frac{1}{s} \label{bcqm1} \\
&&\tilde q_-(z \to -\infty,s) = 0 \;. \label{bcqm2}
\end{eqnarray} 
Identifying the equation for $\tilde q_-(z,s)$ (\ref{qtilde-hyper-G}) as a standard hypergeometric differential equation, its general solution can be written as

\begin{align}
&\tilde{q}_-(z,s)=C_1~F(a',b',c'_-,z) \label{sol1}\\
&+~C_2~z^{1-c'_-}~F(1+a'-c'_-,1+b'-c'_-,2-c'_-,z),   \nonumber 
\end{align}
where $F(a',b',c',z) =  _2\!\!F_1(a',b',c',z)$ is a standard hypergeometric function (the same as before). 
The constants $C_{1,2}$ are fixed from the two boundary conditions (\ref{bcqm1}) and (\ref{bcqm2}). Simplifying, we finally get
\begin{align}
&\tilde{q}_-(z,s)= -\frac{1}{s}\frac{F\left( \frac{s}{\mu}, \frac{s+2\gamma}{\mu}, \frac{s+\gamma}{\mu},z \right)}{F\left( \frac{s}{\mu}, \frac{s+2\gamma}{\mu}, \frac{s+\gamma}{\mu},\frac{1}{2} \right)}. 
\label{qtilde-sol_-0}
\end{align}
From Eq. (\ref{eqnqtilde-harm-2}), we can then obtain the solution for $\tilde q_+(z,s)$. Using some identities of hypergeometric functions, one can simplify and obtain the following result
\begin{align}
\label{qtilde-sol_+0}
&\tilde{q}_+(z,s)= 
-\frac{\gamma}{s} \frac{F\left( \frac{s}{\mu}, \frac{s+2\gamma}{\mu}, \frac{s+\gamma+\mu}{\mu},z \right)}{\mathcal{D}(s,\gamma,\mu)} \end{align}
where the denominator $\mathcal{D}(s,\gamma,\mu)$ is given by
\begin{eqnarray}
&&\mathcal{D}(s,\gamma,\mu)=(\gamma+s)~F\left( \frac{s}{\mu}, \frac{s+2\gamma}{\mu}, \frac{s+\gamma+\mu}{\mu},\frac{1}{2}\right) \label{deno-s} \\
&&+\frac{1}{2}\frac{s(s+2\gamma)}{(s+\gamma+\mu)}~F\left(\frac{s+\mu}{\mu},\frac{s+2\gamma+\mu}{\mu},\frac{s+\gamma+2\mu}{\mu},\frac{1}{2}\right)\;. \nonumber 
\end{eqnarray}
{Using Eq.~\eqref{shift}, we can then express the Laplace transform of the survival probability $\tilde{Q}_-(x_0,s)$ as
\begin{equation}
\tilde{Q}_- (x_0,s) =\frac{1}{s} \left[1-
\frac{F\left( \frac{s}{\mu}, \frac{s+2\gamma}{\mu}, \frac{s+\gamma}{\mu},\frac{1}{2}(1-\frac{\mu x_0}{v_0}\right)}{F\left( \frac{s}{\mu}, \frac{s+2\gamma}{\mu}, \frac{s+\gamma}{\mu},\frac{1}{2} \right)}.
\right] 
\label{qtilde-sol_-}
\end{equation}  
Similarly, $\tilde{Q}_+(x_0,s)$ can be expressed as
\begin{equation}
\label{qtilde-sol_+}
\tilde{Q}_+(x_0,s)=
\frac{\gamma}{s}\left[  1-\frac{F\left( \frac{s}{\mu}, \frac{s+2\gamma}{\mu}, \frac{s+\gamma+\mu}{\mu},\frac{1}{2}(1-\frac{\mu x_0}{v_0} \right)}{\mathcal{D}(s,\gamma,\mu)} \right]
\end{equation}
}

Inverting these Laplace transforms in Eqs. (\ref{qtilde-sol_-}) and (\ref{qtilde-sol_+}) and obtaining explicitly the survival probabilities $Q_\pm(x_0,t)$ from Eq. (\ref{shift}) at any arbitrary time $t$, seems to be extremely hard. However, one can make progress at late times. Since the particle moves in a confining potential, we would expect that the survival probability $Q(x_0,t)$ decays exponentially in time for large $t$
\begin{eqnarray}\label{exp_decay}
Q(x_0,t) \sim e^{-\theta\,t}
\end{eqnarray}   
with a decay exponent $\theta$. For an active particle, with symmetric initial conditions, we anticipate that both $Q_\pm(x_0,t)$ decay exponentially with the same exponent $\theta$, i.e. $Q_{\pm}(x_0,t) \approx e^{-\theta\,t}$. For a passive Ornstein-Uhlenbeck particle (corresponding to the harmonic potential $V(x) = (\mu/2)\,x^2$) , we recall that the corresponding decay exponent is given simply by
\begin{eqnarray}\label{theta_OU}
\theta_{\rm OU} = \mu \;.
\end{eqnarray}
In this case, it simply coincides with the inverse relaxation time $\mu$, as discussed in the previous section. However, for an active particle, we show below that the decay exponent $\theta$ associated to the survival probability (or equivalently for the first-passage probability) is highly non-trivial and different from the inverse relaxation time discussed in the previous section [see Eq. (\ref{res_lambda0})]. 

To extract the decay exponent $\theta$ from the Laplace transforms $\tilde q_{\pm}(z,s)$ in Eqs. (\ref{qtilde-sol_+}) and (\ref{qtilde-sol_-}), we again note that the exponential decay in real time corresponds to having a pole on the negative $s$-axis in the Laplace space. Hence, from Eq. (\ref{qtilde-sol_-}), the denominator must vanish at $s = -\theta$. Hence the exponent $\theta$ is given by the lowest positive root of   
\begin{eqnarray}\label{root1}
F\left( \frac{-\theta}{\mu}, \frac{2\gamma-\theta}{\mu}, \frac{\gamma-\theta}{\mu},\frac{1}{2} \right)=0 \;.
\end{eqnarray}
Similarly, setting the denominator in Eq. (\ref{qtilde-sol_+}) to 0 gives another equation for $\theta$, $\mathcal{D}(-\theta,\gamma,\mu)=0$, which can be shown to coincide with Eq. (\ref{root1}), upon using certain identities of hypergeometric functions. Clearly, from Eq. (\ref{root1}), the solution $\theta$ can be expressed in a scaling form
\begin{equation}\label{defU}
\theta = \mu~U\left( \frac{\gamma}{\mu} \right).
\end{equation}
where the scaling function $U(x)$ can be evaluated numerically, e.g. with Mathematica (as shown by the solid line in Fig. \ref{fig:theta-harm}). It is easy to derive the asymptotic behaviours of the scaling function $U(x)$. Consider first the limit $\mu \to \infty$ (which corresponds to $x\to 0$ limit of $U(x)$ in Eq. (\ref{defU})). 
In this case the position of the particle gets slaved to the noise $v_0 \sigma(t)$. To see this, we note the Langevin equation $dx/dt = - \mu \,x + v_0 \sigma(t)$ indicates that, as $\mu \to \infty$, one can ignore the term $dx/dt$ and the solution is effectively $x(t) \approx (v_0/\mu)\,\sigma(t)$. The noise field $\sigma(t)$ is just a Poisson process with rate $\gamma$ and hence the probability that it does not change sign up to time $t$ is simply $\sim e^{-\gamma\, t}$. This indicates that $\theta \to \gamma$ as $\mu \to \infty$. Hence the scaling function in (\ref{defU}) must behave as $U(x) \approx x$ as $x \to 0$. In the opposite limit $x \to \infty$, which corresponds to $\gamma \to \infty$, the noise $\sigma(t)$ flips sign extremely rapidly and the process $x(t)$ reduces to the passive Ornstein-Uhlenbeck process for which $\theta = \mu$, as discussed earlier. Hence, we expect $U(x) \to 1$ as $x \to \infty$. Summarising, 
\begin{align}\label{asympt}
U(x) = 
\begin{cases}
x & \text{for} \;\;~x \to 0 \\
1 & \text{for} \;\;~x \to \infty \;.
\end{cases}
\end{align}
We also performed numerical simulations for $\mu = 1$ (and setting $v_0=1$) and determined $\theta$ as a function of $\gamma$, as shown by the filled black dots in Fig. \ref{fig:theta-harm}. The results of the simulations are in excellent agreement with the theoretical result shown by the solid line.

The exponential decay of the survival probability $Q(x_0,t) \sim e^{-\theta t}$ for large $t$ in Eq. (\ref{exp_decay}) indicates that the first-passage probability density also decays exponentially with the same rate, $F_{\rm fp}(t,x_0) = -\partial_t Q(x_0,t) \sim e^{-\theta t}$, where $\theta$ is given by the lowest positive root of Eq.~(\ref{root1}) and plotted as a function of $\gamma$ in Fig. \ref{fig:theta-harm}. Indeed the mean first-passage time, which is finite, was already computed exactly in Ref. \cite{Lindenberg}, however the full first-passage time distribution was not studied in detail.

{We note that for the harmonic potential,  the leading inverse relaxation time $\lambda_0=\mu$ starting from a generic initial point $x_0\ne 0$ in  Eq.~\eqref{res_lambda0} and the first-passage exponent $\theta$ in Eq.~\eqref{defU} are different from each other. Indeed for a general $\gamma$, the first-passage exponent $\theta \le \lambda_0=\mu $. One way to think about these two inverse time scales is as follows.  Consider a stochastic Markov process in  a confining potential in presence of an absorbing boundary at $x=a$. When $a\to \infty$, the spectrum of the Fokker-Planck operator will have the lowest eigenvalue $E_0(a\to \infty)=0$ and the first excited state
$E_1(a\to \infty)$ (which is also the gap) gives the inverse relaxation time, i.e., $ \lambda_0= E_1(a\to \infty)$. Now, imagine bringing the absorbing wall from $\infty$ to a finite value $a$. The spectrum of the Fokker-Planck operator will change continuously with $a$.
As $a$ becomes finite, the ground state $E_0(a)$ is precisely the first-passage exponent $\theta(a)$.  There exists an interlacing
theorem~\cite{Hartich18}  that predicts, among other things, that $E_0(a)\le  E_1(\infty)$, i.e., $\theta(a)\le  \lambda_0$. Our result for the active RTP in a harmonic well for $a=0$, namely $\theta(0)\le \lambda_0$ seems to also satisfy this inequality.}
 
\vspace*{0.5cm} 
 
\section{Conclusion}\label{sec:conclusion}

{

To summarise, we have studied the static and dynamics of a run and tumble particle (RTP)
in one-dimension in presence of an external confining potential of the type $V(x) = \alpha \, |x|^p$, with $p>0$.
{We showed that the stationary probability density $P(x)$ has a rich behavior in the $(p, \alpha)$-plane. For $p>1$, the distribution has a finite support in $[x_-,x_+]$ and 
there is an interesting shape transition in
the $(p, \alpha)$-plane across a critical line $\alpha_c(p)$.
For $\alpha > \alpha_c(p)$, the distribution $P(x)$ diverges at $x_\pm$, characteristic of an active-like phase while for $\alpha < \alpha_c(p)$, $P(x)$ 
vanishes at $x_\pm$  corresponding to a  passive-like phase.}
On the marginal line $p=1$, there is an additional
transition as a function of $\alpha$: for $\alpha < \alpha_c = v_0$, the stationary distribution $P(x)$ is
a symmetric exponential, while for $\alpha \geq \alpha_c = v_0$, the stationary distribution collapses 
to a delta function at the origin, $P(x) = \delta(x)$. Furthermore, for $0<p<1$ and all $\alpha > 0$, the stationary
state is again a delta function $P(x) = \delta(x)$. These results are summarised in the phase diagram in Fig. \ref{Fig_Ph_Diag}.
{In addition, for the harmonic case $p=2$, we have obtained exact results for the relaxation to the stationary state and also the probability
of the first-passage time to the origin. 
There is another exactly solvable case for the relaxation as well as the first passage properties.  This corresponds to the marginal case $p=1$, i.e., the active particle moving in a potential  $V(x) = \alpha \, |x|$. The first-passage properties of this model to {an absorbing boundary at an arbitrary position $a$} turns out to be extremely interesting. These results will be reported in a forthcoming  publication.   
}

There remain several interesting open questions. While for the two special cases $p=2$ (harmonic potential) and $p=1$ (marginal case) 
we have obtained exact relaxation and the first-passage properties, 
it remains challenging to obtain analytical results for other values of $p$. Another
interesting extension of our results would be to investigate the relaxation and first-passage properties of an RTP in
higher dimensions. In higher dimensions, there exist some approximate results for the statics and dynamics 
of an RTP in a confining potential \cite{fox1986uniform,Hanggi95,Fodor2016}. However, first-passage properties
have not been systematically studied for an RTP in a confining potential in higher dimensions. 
It would be interesting to find an exactly solvable model in higher dimensions, both for the relaxation to the steady-state
as well as for the first-passage properties. Very recently, the free RTP in one-dimension but in the presence of resetting to the 
initial position was exactly solved both for the stationary state as well as for the first-passage properties \cite{rtp_reset}: it would be interesting to see
how an external potential generalises these results.

Finally, we have studied here the statics, dynamics and first-passage properties of a 
run and tumble particle in a confining potential. Another well studied model for active particles is the 
so-called active Brownian motion (ABM) in two-dimensions in the presence of a confining potential.  
While some results on the steady-state properties have been derived \cite{bechinger_active_2016,Sevilla,Potosky2012,Urna2018,dauchot}, 
there are hardly any study on the relaxational dynamics to the steady states and first-passage properties (see however Ref. \cite{Urna2018} for some recent analytical results). It would thus be interesting to find an exactly solvable model for the relaxation and for the first-passage properties.

\begin{acknowledgements}
The authors would like to acknowledge the support from the Indo-French Centre for the promotion of advanced research (IFCPAR) under Project No. 5604-2. SNM acknowledges the support from the Science and Engineering Research Board (SERB, government of India), under the VAJRA faculty scheme (Ref. VJR/2017/000110) during a visit to Raman Research Institute, where part of this work was carried out.
\end{acknowledgements}

\appendix

\begin{widetext}

\section{Derivation of $\tilde{P}(z,s)$ in Eqs.~\eqref{Pzs.0} and \eqref{A-s}} 

\label{appendixA}

{

In this appendix, we derive the solution for $\tilde{P}(z,s)$ given in Eqs.~\eqref{Pzs.0} and \eqref{A-s} of the main text.  The general solution of the second order differential equation \eqref{p-pm-2nd-order-diffeq} is given as a linear combination of independent hypergeometric functions as
\begin{align}
\tilde{P}_+(z,\bar{s})=&A \, F(1-\bar{s},1-2\bar{\gamma}-\bar{s},2-\bar{\gamma}-\bar{s},z) 
+~B~z^{\bar{\gamma}+\bar{s}-1}~F(\bar{\gamma},-\bar{\gamma},\bar{\gamma}+\bar{s},z), \label{tildeP+sol} \\
\tilde{P}_-(z,\bar{s})=&C ~F(1-\bar{s},1-2\bar{\gamma}-\bar{s},1-\bar{\gamma}-\bar{s},z)
+~D~z^{\bar{\gamma}+\bar{s}}~F(1+\bar{\gamma},1-\bar{\gamma},1+\bar{\gamma}+\bar{s},z), \label{tildeP-sol} 
\end{align}
where {$F(a,b,c,z) \equiv \, _2\,\!F_1(a,b;c; z)$ }is a standard hypergeometric function and $A, B, C$ and $D$ are constants yet to be fixed. We first note that 
the constants $C$ and $D$ are related to $A$ and $B$. This is because $\tilde{P}_+$ and $\tilde{P}_-$ satisfy coupled first-order differential equations. Indeed, by plugging these general solutions in the two first-order differential equations \eqref{eqn:P2-harmonic-laplace-z}, for $z<1/2$. It is then easy to see that  
\begin{equation}
C= - \frac{1-\bar{\gamma}-\bar{s}}{\bar{\gamma}}~A~~~\text{and}~~
D= \frac{\gamma}{\bar{\gamma}+\bar{s}}~B \;.  \label{constants-rela}
\end{equation}
Consequently the solutions then read
\begin{align}
&\tilde{P}_+(z,\bar{s})=A~F(1-\bar{s},1-2\bar{\gamma}-\bar{s},2-\bar{\gamma}-\bar{s},z) 
+~B~z^{\bar{\gamma}+\bar{s}-1}~F(\bar{\gamma},-\bar{\gamma},\bar{\gamma}+\bar{s},z), \label{tildeP+sol-1} \\
&\tilde{P}_-(z,\bar{s})=A~\frac{\bar{\gamma}+\bar{s}-1}{\bar{\gamma}} ~F(1-\bar{s},1-2\bar{\gamma}-\bar{s},1-\bar{\gamma}-\bar{s},z) 
+~B~\frac{\gamma}{\bar{\gamma}+\bar{s}}~z^{\bar{\gamma}+\bar{s}}~ F(1+\bar{\gamma},1-\bar{\gamma},1+\bar{\gamma}+\bar{s},z). \label{tildeP-sol-1} 
\end{align}
where there are only two constants $A$ and $B$ to be determined. One can check that, the above solutions also satisfy the second equation in \eqref{eqn:P2-harmonic-laplace-z}.

The  two constants $A$ and $B$  can be determined from (i) the jump condition \eqref{discont-cond-single} and (ii) the normalisation condition {in Eq. (\ref{normalisation_3}) with $v_0 = 1$. Using the symmetry condition across $z=1/2$, this normalisation condition in Eq. (\ref{normalisation_3}) reduces to 
\begin{align}
\int_0^{\frac{1}{2}}[\tilde{P}_+(z,\bar{s})+\tilde{P}_-(z,\bar{s})] dz  
= \frac{1}{4\bar{s}}. 
\label{normalization}
\end{align}}
Fixing the two constants $A$ and $B$ by the two above conditions (i) and (ii), after straightforward algebra, we find  that $\tilde{P}(z,\bar{s})=\tilde{P}_+(z,\bar{s})+\tilde{P}_-(z,\bar{s})$, for $z<1/2$, reads
\begin{align}
A=0,~~
B=\frac{2^{\bar{\gamma}+\bar{s}-2}}{\,- F\left(1-\bar{\gamma},\bar{\gamma};\bar{\gamma}+\bar{s};\frac{1}{2}\right)+2 \, F\left(-\bar{\gamma},\bar{\gamma};\bar{\gamma}+\bar{s};\frac{1}{2}\right)}. 
   \label{B}
\end{align}

The denominator in the expression for $B$ in  Eq.~\eqref{B} can further be simplified using the identities of hypergeometric function giving
\begin{eqnarray}
 -F\left(1-\bar{\gamma} ,\bar{\gamma}, \bar{s}+\bar{\gamma}, \frac{1}{2}\right)+2 \, F\left(-\bar{\gamma} ,\bar{\gamma}, \bar{\gamma}+\bar{s}, \frac{1}{2}\right)
=\frac{\sqrt{\pi}\,2^{1-\bar{\gamma}-\bar{s}}\,\Gamma[\bar{\gamma}+\bar{s}]}{
\Gamma \left[ \frac{\bar{s}}{2}\right] \Gamma\left[ \frac{1+2\bar{\gamma}+\bar{s}}{2}\right]} \;. \label{deno-app}
\end{eqnarray}
Substituting $A$ and $B$ in Eqs.~\eqref{tildeP+sol-1} and \eqref{tildeP-sol-1}, and adding them gives Eqs
Eqs.~\eqref{Pzs.0} and \eqref{A-s} in the main text.

}

\end{widetext}

\end{document}